%% file: PLAIN_TEXT_arxiv.tex
\documentclass[11pt]{article}
\usepackage{amsmath}
\usepackage{graphicx,psfrag,epsf}
\usepackage{enumerate}
\usepackage{natbib}
\usepackage{url} % not crucial - just used below for the URL 
\usepackage[margin=1in]{geometry}
\usepackage{amsfonts}
\usepackage{algorithm}
\usepackage{algpseudocode}
\usepackage{dsfont}
\usepackage[T1]{fontenc}
\newtheorem{theorem}{Theorem}
\newtheorem{definition}{Definition}
\newtheorem{assumption}{Assumption}
\newtheorem{proposition}{Proposition}[section]
\usepackage[font=footnotesize]{caption}
\newcommand{\pr}{\text{P}}

\newcommand{\E}{\mathbb{E}}

\newcommand{\bX}{\mathbf{X}}
\newcommand{\bx}{\mathbf{x}}

\newcommand{\R}{{\mathds R}}

\newcommand{\Norm}{\mathcal{N}}
\newcommand{\iv}{\mathbb{I}} %indicator variable
\renewcommand{\P}{\mathds{P}} %pro
\newcommand{\PM}{\textsc{pm}$_{2.5}$ }
\newcommand{\PMns}{\textsc{pm}$_{2.5}$}
\newcommand{\sun}{\textsc{SUN}}

\newcommand{\indep}{\perp \!\!\! \perp}

\usepackage[table, dvipsnames]{xcolor} 
\usepackage[resetlabels]{multibib}
\newcites{App}{References}
\usepackage{hyperref}
\hypersetup{
colorlinks=true,
citecolor=blue,
linkcolor=black,
filecolor=blue,      
urlcolor=black,
}
\definecolor{lightGray}{HTML}{DEDEDE}
\definecolor{Gray2}{HTML}{F7F7F7}
\def\percentomila{\ensuremath{{}^\text{o}\mkern-5mu/\mkern-3mu_\text{oooo}}}

\begin{document}

\def\spacingset#1{\renewcommand{\baselinestretch}%
{#1}\small\normalsize} \spacingset{1}

%%%%%%%%%%%%%%%%%%%%%%%%%%%%%%%%%%%%%%%%%%%%%%%%%%%%%%%%%%%%%%%%%%%%%%%%%%%%%%

{
  \title{\LARGE \bf  Bayesian Nonparametrics for Principal Stratification with Continuous Post-Treatment Variables
  %: an Application on Environmental Policies Effects on Health
  }
   \author{Dafne Zorzetto\\
 Data Science Institute, Brown University, Rhode Island, USA \vspace{.2cm}\\ 
 Antonio Canale \\
 Department of Statistics, University of Padova, Italy \vspace{.2cm}\\
 Fabrizia Mealli \\
 Department of Economics, European University Institute, Italy \vspace{.2cm}\\
 Francesca Dominici \vspace{.1cm}\\ 
 Department of Biostatistics, Harvard School of Public Health, Massachusetts, USA \vspace{.3cm}\\ 
 Falco J. Bargagli-Stoffi \vspace{.1cm}\\
Department of Biostatistics, University of California, Los Angeles\\
\texttt{falco@g.ucla.edu}}

\date{} 
\maketitle

\begin{abstract}
Principal stratification provides a causal inference framework for investigating treatment effects in the presence of a post-treatment variable. Principal strata play a key role in characterizing the treatment effect by identifying groups of units with the same or similar values for the potential post-treatment variable at all treatment levels. The literature has focused mainly on binary post-treatment variables. Few papers considered continuous post-treatment variables. In the presence of a continuous post-treatment, a challenge is how to identify and characterize meaningful coarsening of the latent principal strata that lead to interpretable principal causal effects. This paper introduces the Confounders-Aware SHared atoms BAyesian mixture (CASBAH), a novel approach for principal stratification with binary treatment and continuous post-treatment variables. CASBAH leverages Bayesian nonparametric priors with an innovative hierarchical structure for the potential post-treatment outcomes that overcomes some of the limitations of previous works. Specifically, the novel features of our method allow for (i) identifying coarsened principal strata through a data-adaptive approach and (ii) providing a comprehensive quantification of the uncertainty surrounding stratum membership. Through Monte Carlo simulations, we show that the proposed methodology performs better than existing methods in characterizing the principal strata and estimating principal effects of the treatment. Finally, CASBAH is applied to a case study in which we estimate the causal effects of US national air quality regulations on pollution levels and health outcomes.
\end{abstract}

\noindent%
{\it Bayesian causal inference, dependent Dirichlet process, principal causal effects, shared atom mixture model.}
\vfill

%\newpage
\spacingset{1.9} % DON'T change the spacing!

%\listoftodos

\large

\section{Introduction}
\label{sec: intro}

\input{Sections/1_introduction}

\section{Causal Inference Setup}
\label{sec: princ_strata}
\input{Sections/2_setup}

\section{Model}
\label{sec: model}
\input{Sections/3_model}

\section{Simulation Study}
\label{sec: simulation}
\input{Sections/4_simulation_arxiv}

\section{Assessing the Impact of Environmental Policies on Particulate Pollution and Health}
\label{sec: data}
\input{Sections/5_application}

\section{Discussion}
\label{sec: conclusion}
\input{Sections/6_conclusion}

%-----------------------------------------------------------------------------------------------------
{\small
\bibliographystyle{chicago}
\bibliography{Biblio}
}
%-----------------------------------------------------------------------------------------------------
\pagebreak

\appendix
%\counterwithin{equation}{section}
%\counterwithin{figure}{section}
%\counterwithin{table}{section}
\pagenumbering{arabic}
\setcounter{page}{1}
    
\begin{center}
    \textbf{\large SUPPLEMENTARY MATERIAL TO\\ ``Bayesian Nonparametrics for Principal Stratification with Continuous Post-Treatment Variables''}\\ \vspace{0.25cm}
    \normalsize DAFNE ZORZETTO, ANTONIO CANALE, FABRIZIA MEALLI, FRANCESCA DOMINICI, AND FALCO J. BARGAGLI-STOFFI
\end{center}

\section{Conjugate prior distribution for probit regression parameters}
\input{Appendix/appendix_sun_dist}

%\section{Proof Theorem 1}
%\label{app:proof_estimand}
%\input{Appendix/proof_estimand}

\section{Proof Dissociative Stratum Probability}
\label{app:proof_diss_prob}
\input{Appendix/proof_diss_prob}

\section{Posterior Computation}
\label{sec:app_gibbs}
\input{Appendix/appendix_Gibbs}

\section{More simulations details}
\label{sec:app_sim}
\input{Appendix/appendix_simulations}

\section{More application details and results}
\label{sec:app_aplication}
\input{Appendix/appendix_application}

\end{document}

%% file: Sections/1_introduction.tex
Principal stratification \citep{frangakis2002principal} provides a useful framework for estimating causal effects and investigating causal mechanisms in the presence of post-treatment variables (also referred to as intermediate variables). This framework has become a cornerstone of modern causal inference for problems involving noncompliance, truncation by death, and surrogate endpoints. Central to principal stratification are the concepts of \textit{principal strata} and the associated \textit{principal causal effects}. Units are grouped into \textit{principal strata} according to the relationship between the potential outcomes for the intermediate variable under different levels of the treatment. The principal strata are independent of treatment assignment and, thus, represent an inherent latent characteristic of the units. Within each stratum, comparisons of potential primary outcomes across treatment levels define well-posed causal estimands, known as \textit{principal causal effects}. In particular, associative and dissociative effects characterize and quantify treatment effects for units whose intermediate responses would change or not under treatment \citep{frangakis2002principal, mealli2012refreshing}.

Most methodological developments in principal stratification have focused on settings in which both treatment and the post-treatment variable are binary \citep[e.g.,][]{angrist1996identification}. In this case, the number of principal strata is at most four facilitating both interpretation and modeling \citep[see, among many others,][]{cheng2006bounds, imai2008sharp, ding2011identifiability, mattei2011augmented, mealli2013using, mealli2016identification, jiang2016principal, jiang2022multiply, mattei2024assessing}. When the post-treatment variable is continuous, however, the framework induces a potentially infinite collection of principal strata, indexed by the joint potential intermediate responses under treatment and control. This feature fundamentally alters the problem: the strata are latent objects with no natural discretization, rendering classical stratification strategies impractical and complicating both identification and interpretation.

Existing approaches in principal stratification address this difficulty by imposing additional structure. A common strategy is to dichotomize the continuous post-treatment variable \citep{sjolander2009sensitivity, jiang2022multiply}, thereby reducing the problem back to a finite number of strata. While appealing for its simplicity, dichotomization requires selecting thresholds that are often arbitrary and may affect the conclusions \citep{zigler2012estimating, antonelli2023principal}. Alternatively, one may specify fully parametric or semiparametric models for the post-treatment variable conditional on treatment \citep{jin2008principal, conlon2014surrogacy, lu2023principal, schwartz2011bayesian}, or adopt nonparametric modeling strategies \citep{antonelli2023principal}. Although these approaches provide valuable tools, they typically rely on restrictive structural assumptions or require externally imposed criteria to coarsen the infinite set of latent strata. As a result, the analyst must choose between arbitrariness and rigidity.

\subsection{Contributions}

In this work, we propose a Bayesian framework that rethinks principal stratification for continuous post-treatment variables. Rather than fixing a priori strata through discretization or parametric assumptions, we treat principal strata as latent subpopulations with common expected values for the post-treatment variable. To implement this idea, we introduce a Confounders-Aware SHared-atoms BAyesian mixture model (CASBAH), built upon a dependent Dirichlet process prior \citep{maceachern1999dependent, quintana2020dependent}. CASBAH flexibly models the distribution of the potential post-treatment variable under each treatment level, conditional on observed confounders, while employing shared mixture atoms across treatments. Under this formulation, the existence of dissociative stratum has non-null prior probability. This feature is particularly appealing from a causal perspective: it allows, in a probabilistically coherent manner, the possibility that such strata exist in the population, rather than requiring their presence to be imposed deterministically or ruled out a priori as a byproduct of modeling choices.

Our proposed approach advances the literature in several ways. First, by incorporating confounders directly into the mixture weights (similarly to \citep{zorzetto2024confounder}), CASBAH enables flexible adjustment for measured confounding within a fully nonparametric specification of the intermediate response distribution. Second, the shared-atoms formulation allows the model to identify coarsened associative and dissociative strata endogenously, eliminating the need for arbitrary thresholding or externally imposed clustering rules. Third, the framework jointly imputes missing potential intermediate responses and outcomes, and propagates uncertainty in principal strata membership through posterior inference, thereby providing coherent uncertainty quantification for principal causal effects. Together, these features yield a flexible and interpretable approach to principal stratification in settings where the intermediate variable is continuous and traditional methods are inadequate.

%% file: Sections/2_setup.tex
We assume that we observe $n$ independent units. For each unit $i\in \{1, \dots, n\}$, let $\bX_i \in \mathcal{X} \subseteq \mathbb{R}^p$ the set of observed covariates, $T_i \in \{0,1\}$ the observed (binary) treatment, $P_i \in \mathcal{P} \subseteq \mathbb{R}$ the post-treatment variable, and $Y_i \in \mathcal{Y} \subseteq \mathbb{R}$ the primary response. For each unit $i$, the potential post-treatment outcomes are defined as $\{P_i(0), P_i(1)\} \in \R^2$ and the potential primary outcomes are defined as $\{Y_i(0), Y_i(1)\} \in \R^2$, for $i = 1, \dots, n$. The vector $\{P_i(0), P_i(1)\}$ represents the collection of the two potential values of the post-treatment variable,  when the unit $i$ is assigned to the control or the treatment group, respectively. Similarly, the vector $\{Y_i(0), Y_i(1)\}$ represents the collection of the two potential values for the response variable, with $Y_i(0)$ when the unit $i$ is assigned to the control group and $Y_i(1)$ when assigned to the treatment group.

%\subsection{Identifying Assumptions}
%\label{sec:setup}

Following the principal stratification framework \citep{frangakis2002principal, vanderweele2011principal, mealli2012refreshing, feller2017principal, ding2017principal, lu2023principal}, the principal strata are not affected by treatment assignment; therefore, a principal stratification can be used as any unit classification to define meaningful causal estimates conditional on the principal strata and discover the heterogeneous treatment effect \citep{mealli2012refreshing}.

As highlighted in the Introduction, the definition of the \textit{dissociative} and \textit{associative} strata is of particular interest. Formally, a dissociative stratum includes units where the treatment does not affect the post-treatment variable, that is, $P_i(0)$ is the same as $P_i(1)$. In contrast, the \textit{associative stratum} includes units for which the treatment affects the post-treatment variable, yielding a positive effect, \textit{positive associative stratum}, or a negative effect, \textit{negative associative stratum}. 

Conditional on principal strata, the estimands of interest are the \textit{principal causal effects} (PCE), such that
\begin{equation}
    \E\{Y_i(1)-Y_i(0) \mid g\{P_i(1),P_i(0)\} \in \mathcal{S}_s\},\label{eq:casual_estimand}
\end{equation}
where $g(\cdot)$ is a functional of the potential post-treatment values and $\mathcal{S}_s \subseteq \mathbb{R}$ indicates the subset of values that identify each principal stratum $s$. 

Throughout the paper, we invoke the following assumptions:

\begin{assumption}[Stable Unit Treatment Value Assumption]\label{ass:sutva}
 \begin{eqnarray*} 
 & Y_i(T_1, T_2, \cdots, T_i, \cdots,  T_n) =  Y_i(T_i),  \quad Y_i(T_i) = Y_i, \mbox{  for } i=1,\dots,n;\\
& P_i(T_1, T_2, \cdots, T_i, \cdots,  T_n) = P_i(T_i),  \quad P_i(T_i)=P_i, \mbox{  for } i=1,\dots,n.
\end{eqnarray*} 
\end{assumption}

The Stable Unit Treatment Value Assumption is a combination of (i) no interference between units, that is, the potential outcome and the potential values of the post-treatment variable of the unit $i$ do not depend on the treatment applied to other units, and (ii) consistency, that is, no different versions of the treatment levels assigned to each unit \citep{rubin1986comment}. Under the principal stratification framework, this assumption is invoked for both the primary outcome variable and the post-treatment variable. 

In practice, for $i=1,\dots,n$, we observe $p_i \in \R$ and $y_i \in \R$, that is, the realization of the random variables $P_i$ and $Y_i$, respectively, and defined as $P_i = (1-T_i) \cdot P_i(0) + T_i \cdot P_i(1)$ and $Y_i = (1-T_i) \cdot Y_i(0) + T_i \cdot Y_i(1)$.

\begin{assumption}[Strongly Ignorable Treatment Assignment]\label{ass:sita} Given the observed covariates $\bx_i$,
\begin{eqnarray*}
     \{Y_i(1), Y_i(0), P_i(0), P_i(1) \} \indep  T_i \mid X_i,
\end{eqnarray*}
\begin{equation*}
 0 < \pr\left(T_i=1 \mid X_i = x\right) < 1 \:\: \forall \: x \in \mathcal{X}.\nonumber
\end{equation*}
%where $\mathcal{X}$ is the features' space.
\end{assumption}
Assumption~\ref{ass:sita} states that: (i) for each unit, the potential primary outcomes and the potential post-treatment variables are independent of the treatment conditional on the set of covariates $X_i$; (ii) all units have a positive chance of receiving the treatment.

\subsection{Causal Estimands}
\label{subsec:estimand}

Following the general definition of causal estimands in eq.~\eqref{eq:casual_estimand}, we assume that the functional $g(\cdot)$ is the expected value of the difference of the random variable of potential post-treatment for each unit $i$ given the confounders. The partition of $\mathbb{R}$ defining the principal strata is $\mathbb{R} = \bigcup_{s\in\{0,+,-\}} \mathcal{S}_s$,  with $\mathcal{S}_+ = \mathbb{R}_+$, $\mathcal{S}_{-} = \mathbb{R}_-$, and $\mathcal{S}_0 = \{0\}$ for the associative positive stratum, associative negative stratum, and dissociative stratum, respectively. This assumption is formalized in the following definition. 

\begin{definition}[Causal Estimands] \label{eq:ede}

We define the causal estimands in eq.~\eqref{eq:casual_estimand} for the expected dissociative effect ($\tau_0$) and the expected associative effects positive ($\tau_{+}$) and negative ($\tau_{-}$) as 
\begin{align*}
    & \tau_0 = \E\{Y_i(1)-Y_i(0) \mid \E_i\{P_i(1)- P_i(0)\} = 0\}, \\
    &  \tau_{+} = \E\{Y_i(1)-Y_i(0) \mid \E_i\{P_i(1)- P_i(0)\} > 0\}, \\
    & \tau_{-} = \E\{Y_i(1)-Y_i(0) \mid \E_i\{P_i(1)- P_i(0)\} < 0\}.
\end{align*}   
\end{definition}
These causal estimands are not directly identifiable from the data due to the counterfactual outcome of the post-treatment variable and the primary outcome. However, they can be weakly identified by invoking Assumptions \eqref{ass:sutva} and \eqref{ass:sita}.

For clarity, $\E_i\{\cdot\}$ denotes the expectation of a random variable for a specific unit $i$, while $\E\{\cdot \mid \mathcal{A}\}$ denotes the conditional expectation taken over units satisfying the condition $\mathcal{A}$.

\vspace{0.25em}
\noindent 
\begin{proposition}
\label{th:theorem_1} Under assumptions \eqref{ass:sutva} - \eqref{ass:sita}, we can rewrite the causal estimands as 
\begin{align*}
    \int_x &\big[ \E\{Y_i \mid T_i=1, g\{P_i(1),P_i(0)\} \in \mathcal{S}_s, X_i =x\}  - \\
    &\; \E\{Y_i \mid T_i=0, g\{P_i(1),P_i(0)\} \in \mathcal{S}_s, X_i =x\} \big] \pr(X_i =x \mid g\{P_i(1),P_i(0)\} \in \mathcal{S}_s) dx;
\end{align*}
where the expected value of the primary outcome can be rewritten as the following
\begin{align}
    & \E\big\{Y_i \mid T_i=t, g\{P_i(1),P_i(0)\} \in \mathcal{S}_s, X_i=x\big\} \notag \\
    & \quad = \int_{p_0 p_1} \E\big\{Y_i \mid T_i=t,  X_i=x, P_i(1)=p_1,P_i(0)=p_0 \big\}\notag \\
    & \quad\quad\quad  \times \pr\big(P_i(1)=p_1,P_i(0)=p_0\mid T_i=t,g\{P_i(1),P_i(0)\} \in \mathcal{S}_s, X_i=x \big)  dp_0 p_1 dx;
    \label{eq:theorem_1}
\end{align}
where inner expectations $\E\big\{Y_i \mid T_i=t,  X_i=x, P_i(1)=p_1,P_i(0)=p_0 \big\}$ are estimated with the outcome model $Y_i \mid T_i,  X_i, P_i(1),P_i(0)$, as well as probability $\mbox{pr}\big(P_i(1)=p_1,P_i(0)=p_0\mid T_i=t,g\{P_i(1),P_i(0)\} \in \mathcal{S}_s, X_i=x \big)$, which is defined by the model for the potential post-treatment variables.
\end{proposition} %The proof is reported in the Supplementary Material.

%% file: Sections/3_model.tex
\subsection{Confounders-Aware Shared-atoms Bayesian Mixture Model}

Following the Bayesian paradigm and invoking De Finetti's theorem \citep{rubin1978bayesian}, the joint probability distribution of the involved variables is unit exchangeable, and it is defined as 
$$\pr(Y(0), Y(1), P(0), P(1),T,
X)=\int_\Theta \prod_{i=1}^n \pr(Y_i(0), Y_i(1), P_i(0), P_i(1),T_i, X_i \mid \theta) \pr(\theta) d\theta, $$ where $\pr(\theta)$ is the prior measure for all the involved parameters $\theta$ that take values in the parameter space $\Theta$. The inner probability can be factorized as
\begin{align}
    \pr( T_i \mid Y_i(0), Y_i(1), P_i(0), P_i(1), X_i, \theta) & \times \pr(Y_i(0), Y_i(1) \mid P_i(0), P_i(1), X_i, \theta) \notag \\
     \times \pr(P_i(0), P_i(1)  \mid X_i, \theta)  & \times \pr( X_i \mid \theta).
    \label{eq:decom_prob}
\end{align}
Invoking Assumption 2, the conditional probability of the treatment variable can be written as $\pr(T_i \mid Y_i(0), Y_i(1), P_i(0), P_i(1), X_i, \theta)=\pr( T_i \mid X_i, \theta)$. Moreover, we replace the conditional distribution of the covariates with the empirical distribution so that $\pr( X_i \mid \theta)=\pr(X_i)$.

The remaining two probabilities in equation \eqref{eq:decom_prob} must be modeled: (i) the distribution of potential primary outcomes conditional on the potential post-treatment variables and covariates, and (ii) the distribution of the potential post-treatment outcomes conditional on the covariates.

We focus our attention on the latter, that is, the distribution of the potential post-treatment outcomes conditional on the covariates for which we propose a novel Bayesian nonparametric approach. As our primary focus lies in modeling the post-treatment variable distribution, for the sake of clarity in our discussion, we employ a parametric model for the conditional response distribution following the settings of \citet{schwartz2011bayesian}.  Note that, although the primary outcome model is assumed to be a linear regression model, it can also be generalized to a more complex and flexible model. 

For modeling the post-treatment variable, we exploit a dependent nonparametric mixture, following the dependent Dirichlet process approach \citep{mac2000dependent, barrientos2012support, quintana2020dependent}. %In particular, our model also shares some similarities with the hierarchical Dirichlet process \citep{teh2004sharing, teh2006hierarchical, teh2009hierarchical} and the recent common atom model of \citet{denti2023common}.
Specifically, we assume for each $i=1,\dots,n$:
\begin{align}
      \{P_{i} \mid \bx_i,t \} &\sim f^{(t)}( \cdot \mid \bx_i), \mbox{ for } t=\{0,1\},\notag \\ 
       f^{(t)}(\cdot \mid \bx_i) &=\int_\Psi {\mathcal K}(\cdot; \psi) dG_{\bx_i}^{(t)}(\psi), \notag \\ 
       \{G_{\bx_i}^{(0)}, G_{\bx_i}^{(1)}\} & \sim \Pi,  
      \label{eq:model1_p2} 
\end{align}
where ${\mathcal K}(\cdot;\psi)$ is a continuous density function, for every $\psi \in \Psi$, and $G_{\bx_i}^{(t)}$ is a random probability measure depending on the confounders $x_i$ associated to an observation assigned to treatment level~$t$. 
%For each treatment level $t=\{0,1\}$, 
The random probability measures $G_{\bx_i}^{(0)}$ and $G_{\bx_i}^{(1)}$ are defined as % characterized by treatment-specific sets of random weights assigned to a common set of random atoms. Specifically,  %induced by the common process $\Pi$. 
%Following the characterization of the , we write:
\begin{equation}
G_{\bx_i}^{(t)}  = \sum_{l \geq 1} \pi_{l}^{(t)}(\bx_i) \delta_{\mathbb{\psi}_{l}},
\label{eq:asdiscreteG_p2}
\end{equation}
where the sequences $\{\pi_{l}^{(t)}(x_i)\}_{l\geq 1}$ for $t=\{0,1\}$ represent infinite sequences of random weights, and $\{\psi_{l}\}_{l\geq 1}$ is an infinite sequence of random kernel's parameters, independent and identically distributed from a base measure $H$, shared among potential post-treatment outcomes under both treatment levels $t$.

The sequences of weights are defined through a stick-breaking representation \citep{sethuraman1994constructive}, 
\begin{align}
    \pi_{l}^{(t)}(\bx_i)  &= \omega_{l}^{(t)}(\bx_i)\prod_{r<l}\{1-\omega_{r}^{(t)}(\bx_i)\},
\label{eq:model_omega_p2}
\end{align}
where $\{\omega_{l}^{(t)}(\bx)\}_{l\geq 1}$ , for $t=\{0,1\}$ are $[0, 1]$-valued independent stochastic processes.

The discrete nature of the random probability measure $G_{\bx_i}^{(t)}$, for $t=\{0,1\}$, allows us to introduce the latent categorical variables $S_{i}^{(t)}$, that describe the cluster allocations for each unit $i \in \{1, \dots, n\}$, whose clusters are defined by heterogeneous values for $P_i(t)$. Assuming $\Pr(S_{i}^{(t)} = l ) = \pi_{l}^{(t)} (\bx_i)$, we can write model in \eqref{eq:model1_p2}, exploiting conditioning on $S_{i}^{(t)}$, as
\begin{equation}
    \{P_{i} | \bx_i,t, \psi, S_{i}^{(t)} = l\}  \sim 
    {\mathcal K}( \cdot \mid \bx_i,\psi_{l}),\quad \psi_{l} \sim H.\notag
\end{equation}
where $\psi$ represents the infinite sequence $\{\psi_{l}\}_{l\geq 1}$.

Among the plethora of dependent non-parametric processes, % \citep[see the recent review by][for a detailed description]{quintana2020dependent}, 
we focus on the probit stick-breaking process %for its good theoretical properties and ease of computation
\citep{rodriguez2011nonparametric} and specifically assuming 
\begin{align}
    \omega_{l}^{(t)}(x_i)  = \Phi(\alpha_{l}^{(t)}(x_i)), \quad 
    \alpha_{l}^{(t)}(x_i) \sim \Norm( \beta_{
    l0}^{(t)} + x_i^T\beta_{l}^{(t)},1),
    \label{eq:model4_p2}
\end{align}
where $\Phi(\cdot)$ is the cumulative distribution function of a standard Gaussian distribution, $\{\alpha_{l}^{(t)}(x_i) \}_{l\geq 1}$ has Gaussian distributions with mean a linear combination of the $p$ covariates $x_i$, and $\{\beta_{l0}^{(t)},\beta_{l}^{(t)}\}_{l\geq 1}$ are the treatment-specific probit regression parameters.
For the regression parameters in \eqref{eq:model4_p2}, we assume the following multivariate Gaussian prior:
\begin{equation}
    %\beta^{(t)} \sim \sun_{(p+1)(L-1),h}(\bxi,\Omega,\Delta,\gamma,\Gamma), 
    \beta^{(t)} \sim \Norm_{(p+1)(L-1)}(\xi,\Omega),
    \label{eq:prior_beta}
\end{equation}
for $t=0,1$ and $l\geq 1$. Consistently with \citet{fasano2022class}, the Gaussian prior leads to a straightforward posterior computation, as discussed in the next section.

We assume the kernel  density to be Gaussian, so that model \eqref{eq:model1_p2}--\eqref{eq:asdiscreteG_p2} becomes
\begin{align}
     \{P_{i}(t)   \mid S_{i}^{(t)}=l, \eta, \sigma \}\sim \Norm (\eta_{l},\sigma_{l}^2).
      \label{eq:model5_p2} 
      \end{align}
where $\eta$ and $\sigma$ represent infinite sequences of location parameters $\{\eta_{l}\}_{l\geq 1}$ and scale parameters $\{\sigma_{l}\}_{l\geq 1}$, respectively, such that $\psi_l=(\eta_l,\sigma_l)$.  

The prior specification is completed by assuming %assuming that for the parameters $\eta_l$ and $\sigma_l$ in \eqref{eq:model5_p2}
\[
 \eta_l \stackrel{iid}{\sim} \Norm(\mu_\eta,\sigma_\eta^2), \mbox{ and } \sigma_l^2 \stackrel{iid}{\sim} \mbox{InvGamma}(\gamma_1,\gamma_2),
\]
where InvGamma($\gamma_1, \gamma_2$) represents the inverse gamma distribution with the shape parameter $\gamma_1 \in \mathbb{R}^+$ and the scale parameter $\gamma_2 \in \mathbb{R}^+$, and mean equal to $\gamma_2/(\gamma_1-1)$ and variance $\gamma_2^2/\{(\gamma_1-1)^2(\gamma_1-2)\}$.

\subsection{Discovery of Principal Strata}

In the causal inference literature, the definition of the dissociative stratum for continuous post-treatment variables remains a central methodological challenge. When the post-treatment variable is modeled as continuous with a diffuse distribution, the probability that an unit belongs to the dissociative stratum is zero.

Excluding the dichotomization method for post-treatment variables \citep{sjolander2009sensitivity, jiang2022multiply}, which the authors of this manuscript find very extreme, alternatives in the literature the literature has proposed defining approximate dissociation via ad hoc thresholding rules \citep{zigler2012estimating}, whereby individuals are assigned to the dissociative stratum if 
$g\{P_i(1),P_i(0)\}$ falls into a user-specified cutoff. 

In contrast, leveraging the discrete support induced by the Dirichlet process prior, our approach preserves the definition of the dissociative stratum as $g\{P_i(1),P_i(0)\} \in \mathcal{S}_0$ and introduces a fully data-adaptive procedure that assigns a priori a positive probability to belong to this stratum.

Bayesian nonparametric mixtures are well-known to excel at capturing cluster structures, and we tailor them to solve this specific challenge in causal inference framework. Specifically, to estimate our causal estimands of interest, we need to characterize the joint distribution of $\{P_{i}(0), P_{i}(1)\}$, through their latent cluster allocation variable $\{S_i^{(0)},S_i^{(1)}\}$, so that we can determine each unit's allocation to a principal strata. This joint distribution cannot be directly observed due to the fundamental problem of causal inference. Nevertheless, our proposed Bayesian nonparametric model allows for robust estimation of this joint distribution through its construction of shared parameters across the two treatment levels. In this section, we illustrate how CASBAH enables the characterization of principal strata in a statistically robust and fully data-adaptive manner. 

%We start from noting that the random probability measure characterization for the post-treatment variable in \eqref{eq:asdiscreteG_p2} enables us to define two distinct mixtures for each treatment level $t \in \{0,1\}$. In particular, the shared random kernel parameters $\{\psi_{l}\}_{l\geq 1}$ in our model induce two important properties. First, the potential post-treatment variables $\{P(1), P(0)\}$ maintain a dependence on covariates $X$ even conditionally, as demonstrated in the posterior distribution in Section \ref{subsec:posterior} and the Supplementary Material. Second, each subject $i$ has a non-zero probability of belonging to the dissociative stratum. This latter property is proven in Theorem \ref{th:theorem_2}. 

We start from noting that the latent categorical variables $\{S_{i}^{(0)}, S_{i}^{(1)}\}$ for each unit $i \in\{1,\ldots,n\}$ serve a dual purpose. First, they determine the allocation probabilities to specific mixture components. Second, they simultaneously define the marginal distributions of the random unit partition based on potential post-treatment outcomes \citep{quintana2006predictive}. This formulation creates a natural bridge between mixture model clustering and principal stratification, allowing us to characterize the heterogeneity of the treatment effect through a principled probabilistic structure. 

\begin{figure}
\begin{center}
%\figuresize{.18}
%\figurebox{20pc}{25pc}{}[Figures/probdissociative.png]
\includegraphics[width=5in]{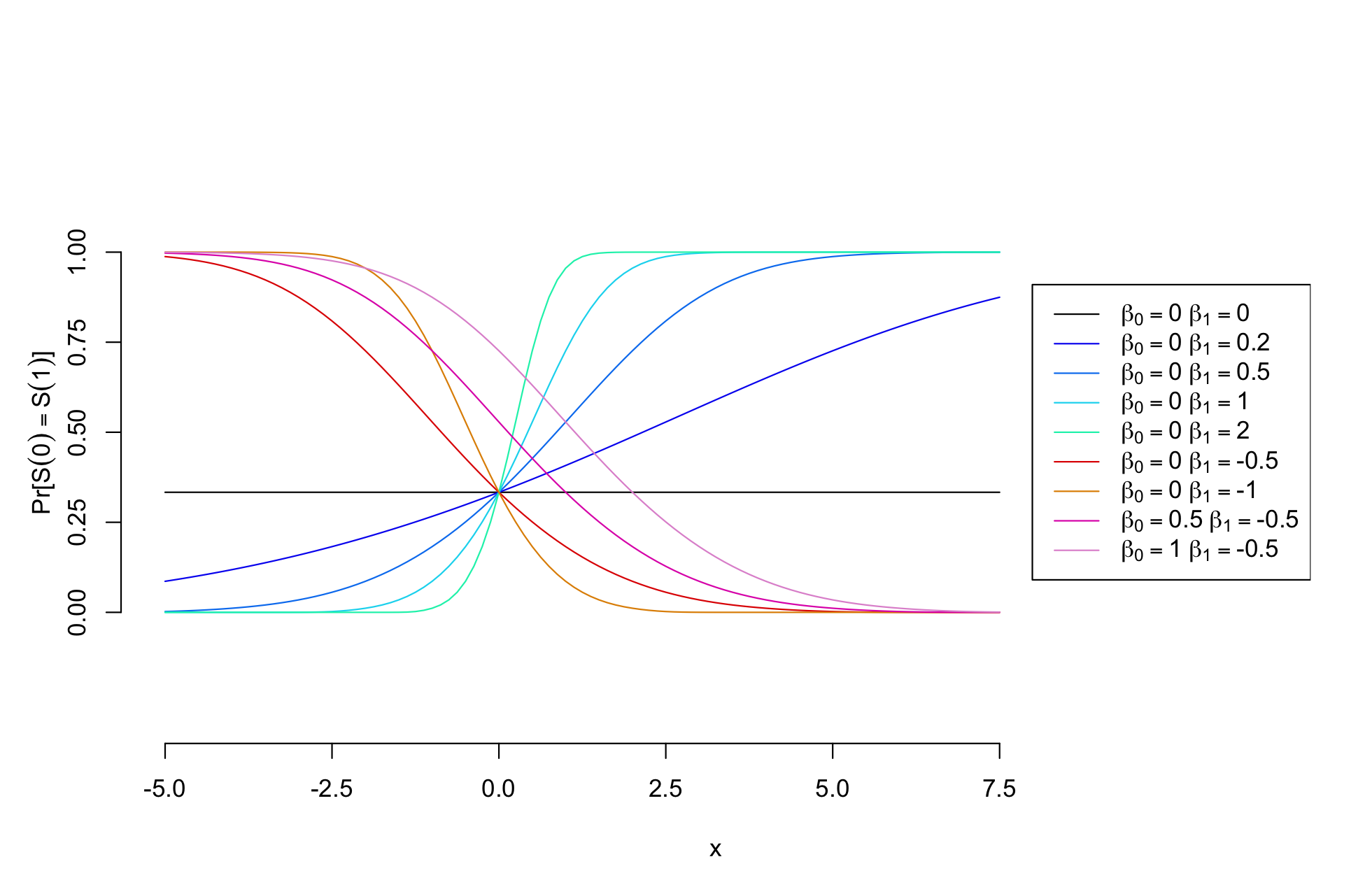}
\end{center}
\caption{Probability to belong to the dissociative stratum. Function varies according to different values of $\alpha(x)=\beta_0+\beta_1x$. \label{fig:prob_diss}}
\end{figure}

CASBAH is defined such that the atoms $\{\psi_l\}_{l \geq 1}$ are shared between the two potential post-treatment outcomes for the same unit. This formulation yields precise posterior probabilities that quantify, for each unit $i$, whether the latent indicators $\{S_{i}^{(0)}, S_{i}^{(1)}\}$ map to identical or distinct components of the mixture. Specifically, for any unit $i \in \{1,\ldots,n\}$, the probability of membership in the dissociative stratum is equals the probability that the atom under control differs from the atom under treatment, i.e. $S_{i}^{(0)} \neq S_{i}^{(1)}$. This probability is provably non-zero and can be derived a priori through the following theorem.
\begin{theorem}
\label{th:theorem_2}  Given the model~\eqref{eq:model1_p2}, the probability a priori to belong to the dissociative stratum is defined as:
\begin{equation}
    \pr\{S^{(0)}=S^{(1)}\} = \frac{\rho_2(x)[\{1+\rho_2(x)-2\rho_1(x)\}^L-1] }{\rho_2(x)-2\rho_1(x)}
    \label{eq:theorem_2}
\end{equation}
where $x$ are the covariates, $\rho_1(x)=\E\big[\Phi\big(\alpha(x)\big)\big]$, $\rho_2(x)=\E\big[\Phi\big(\alpha(x)\big)^2\big]$ and $0 \leq \pr\{S^{(0)}=S^{(1)}\}\leq 1$.
\end{theorem}
The proof is reported in the Supplementary Material. The representation of the probability of belonging to the dissociative stratum in \eqref{eq:theorem_2} is further illustrated in Figure \ref{fig:prob_diss}. 

This probability varies according to the function $\alpha(x)=\beta_0+\beta_1x$. Thus, the probability of belonging to the dissociative stratum depends on the observed covariates $x$ for each unit $i$ and also on the distribution of the parameters $\beta=\{\beta_0, \beta_1\}$.

The posterior distribution of strata allocation is by definition a function of the likelihood such that the probability is driven by the observed data.

\subsection{Posterior Inference}
\label{subsec:posterior}

Theorem \ref{th:theorem_1} establishes that principal causal estimands are characterized as functions of both the primary response and potential post-treatment outcomes, each involving specific model parameters. Following the Bayesian paradigm, we estimate these parameters through their posterior distributions conditional on the observed data. This approach provides not only point estimates but also complete uncertainty quantification.

Sampling from the posterior joint distribution is straightforward via Gibbs sampling. In particular, the proposed algorithm (described in detail in the Supplementary Material) takes inspiration from \citet{teh2004sharing, teh2006hierarchical, teh2009hierarchical} for the hierarchical structure, and from \citet{fasano2022class} for the probit regression in the weights.
We assume a general linear regression for the outcome model that depends on the parameter $\theta_0^Y$ for the outcome under control and $\theta_1^Y$ for the treated outcome.

The implementation of the Gibbs sampler proceeds through the following steps.
\begin{itemize}
    \item The latent variable $S_i^{(t)}$ is a multivariate distribution with
    $$\pr\{S_{i}^{(t)}=l \} \propto \pi_{l}^{(t)}(\bx_i) \pr( P_i(t) \mid S_{i}^{(t)}=l;\eta_{l},\sigma_{l}),$$
    %kokppulypluy l
    where
    $\pi_{l}^{(t)}(\bx_i)=\Phi(\alpha_{l}^{(t)}(\bx_i)) \prod_{r<l} \{1-\Phi(\alpha_{r}^{(t)}(\bx_i))\}$ for $l=1,\dots, L-1$ and $\Phi(\alpha_{L}^{(t)}(\bx_i))=1$.
    \item The cluster-specific parameters are drawn from 
    $$\psi_l \propto \Pi_{x}\prod_{i:S_i^{(0)}=l}\pr(P_i(0) \mid S_i^{(0)}=l, X_i; \psi_l)\prod_{i:S_i^{(1)}=l}\pr(P_i(1) \mid S_i^{(1)}=l, X_i; \psi_l).$$
    \item The logit regression  parameters for conditional-dependent weights are drawn from 
    $$\beta^{(t)} \propto \pr(\beta^{(t)}) \prod_{i=1}^n \pr(P_i(t) \mid S_i^{(t)}, X_i; \beta^{(t)})$$
    following \citep{fasano2022class}. Details are reported in the Supplementary Material.
    \item The missing post-treatment variable is imputed from
    $$P_i(1-T_i) \propto \pr(Y_i \mid P_i(0), P_i(1), X_i; \theta_0^Y)^{1-T_i} \pr(Y_i \mid P_i(0), P_i(1), X_i; \theta_1^Y)^{T_i} \pr(P_i(1-T_i) \mid X_i; \psi). $$
    \item The conditional posterior distribution for the outcome model is built from
    $$\theta_t^Y \sim \pr(\theta_t^Y)\prod_{i=1}^n \pr(Y_i(t) \mid P_i(0), P_i(1), X_i; \theta_t^Y).$$
\end{itemize}

%% file: Sections/4_simulation_arxiv.tex
The performance of the proposed mixture model is evaluated through a simulation study. Our objective is to investigate the model's ability to (i) accurately impute the missing post-treatment and outcome variables, that is, to assess the potential bias of the expected value $E\{P_i(1)-P_i(0)\}$ and $E\{Y_i(1)-Y_i(0)\}$ over the units $i=1\dots,n$, (ii) correctly identify the principal strata and estimate the principal causal effects. To achieve this, we conducted simulations under five different data-generating mechanisms and analyzed the results to understand the model's behavior in different scenarios.

The performance of our proposed approach is compared to those obtained with the \citet{schwartz2011bayesian}'s model and the copula model proposed by \citet{lu2023principal}.

Commonly across the scenarios, we assume a linear regression model for the outcome model, defined as follows:
\begin{equation}
    \begin{bmatrix} Y_i(0) \\ Y_i(1)
         \end{bmatrix} \sim \Norm_2 \begin{pmatrix}
          \begin{bmatrix}
           \theta_{00} + \theta_{01}P_i(0) \\           
           \theta_{10} + \theta_{11}P_i(1)+ \theta_{12}P_i(0) + \theta_{13}P(0)P_i(1)
          \end{bmatrix},
          \begin{bmatrix}
           e^{\lambda_0} & 0 \\
           0 & e^{\lambda_0 +\lambda_1 P_i(1)}
         \end{bmatrix}
    \end{pmatrix}.
\label{eq:Y_model}
\end{equation}
Assuming as prior distribution for the parameters
\begin{align*}
    \theta^{(0)} =(\theta_{00},\theta_{01}) \sim \Norm_2(\mu_{\theta},\sigma^2_{\theta} I_2) & \mbox{ and }
    \theta^{(1)} =(\theta_{10},\theta_{11},\theta_{12},\theta_{13})\sim \Norm_4(\mu_{\theta},\sigma^2_{\theta} I_4);\\
    \lambda_0 \sim \Norm(\mu_\lambda,\sigma^2_{\lambda}) & \mbox{ and } \lambda_1 \sim \Norm(\mu_\lambda,\sigma^2_{\lambda});
\end{align*}
with $\mu_\theta=0$, $\sigma_\theta=10$, $\mu_\lambda=0$, and $\sigma_{\lambda}=2$.  Clearly, CASBAH can accommodate more complex choices for the outcome, but we decided to opt for a simpler model following what had already been done in the literature by \citet{schwartz2011bayesian} and \citet{lu2023principal}. 

For the data generating process, we simulate two Bernoulli confounders $(X_1, X_2)$ for Scenarios 1 to 4 and five Bernoulli confounders $(X_1, X_2, X_3, X_4, X_5)$ for Scenario 5, and a binary treatment variable such that $T_i \sim \mbox{Be}(\mbox{logit}(0.4X_{1i}+0.6X_{2i}))$ for Scenarios $1-4$ and $T_i \sim \mbox{Be}(\mbox{logit}(0.4X_{1i}+0.6X_{2i}-0.3X_{3i}+0.2X_{4i}X_{5i}))$ for Scenario 5, for $i =1, \dots, n$. 

Each scenario assumes a different conformation of the strata for the continuous post-treatment variable $P_i=(P_i(0), P_i(1)) \in \mathbb{R}^2$. Each stratum is obtained by introducing, for each unit, categorical variables $S_{i}^{(0)}$ and $S_{i}^{(1)}$, for control and treatment levels, respectively, with the vector of probabilities that depend on the values of the confounders and allocates the unit in different clusters. Conditionally on the cluster allocation $S_i^{(t)}=s$, with $s$ that has the same support for control and treatment levels, we simulate both potential post-treatment variables, under control and under treatment, respectively, as
\begin{equation*}
   \{ P_i(0)|S_{i}^{(0)}=s\} \sim \Norm(\eta_{s},\sigma^2_{s}), \quad
    \{P_i(1)|S_{i}^{(1)}=s \}\sim \Norm(\eta_{s},\sigma^2_{s}).
\end{equation*}
Continuous potential outcomes $(Y_i(0),Y_i(1))$, for $i=1,\dots,n$, are simulated following the model \eqref{eq:Y_model} with common values for $\lambda_0=-0.5$ and $\lambda_1=0.1$, while $\theta^{(0)}$ and $\theta^{(1)}$ are different for each scenario.
In each setting, the sample size is set to $n=500$. For each scenario, we simulate $100$ samples. Below we provide further details for each one of the five scenarios.

\vspace{0.25cm}
\noindent {\em Scenario 1:} We investigate a situation in which there are two strata: one with a dissociative effect and one with a positive associative effect. In particular, for $S_{i}^{(0)}=S_{i}^{(1)}=1$ $P_i(0), P_i(1) \sim \Norm(1,0.05)$, and for $S_{i}^{(0)}=2$ and $S_{i}^{(1)}=3$ $P_i(0)\sim \Norm(2,0.05)$ and $P_i(1) \sim \Norm(3,0.05)$. The regression parameters for the Y-model are $\theta^{(0)}=(1,2)$ and $\theta^{(1)}=(1,2,-1,0.5)$.

\vspace{0.25cm}
\noindent {\em Scenario 2:} We focus on a case where we have a dissociative stratum, for $S_{i}^{(0)}=S_{i}^{(1)}=1$ both $P(0)$ and $P(1)$ are simulated from $\Norm(2,0.05)$, an associative stratum with a positive effect, for $S_{i}^{(0)}=2$ and $S_{i}^{(1)}=3$ $P_i(0)\sim \Norm(2,0.05)$ and $P_i(1) \sim \Norm(3,0.05)$, and dissociative stratum with a negative effect, for $S_{i}^{(0)}=2$ and $S_{i}^{(1)}=1$ $P_i(0)\sim \Norm(2,0.05)$ and $P_i(1) \sim \Norm(1,0.05)$. The regression parameters for Y-model are $\theta^{(0)}=(1,2)$ and $\theta^{(1)}=(1,1.2,-1,1)$.

\vspace{0.25cm}
\noindent {\em Scenario 3:} This scenario corresponds to Scenario 1 with closer atoms for the strata and different variances. In particular, the dissociative stratum has $(P_i(0)|S_{i}^{(0)}=1) = (P_i(1)|S_{i}^{(1)}=1) \sim \Norm(1.5,0.12)$, and the associative stratum has $(P_i(0)|S_{i}^{(0)}=2)  \sim \Norm(2,0.1)$ and $(P_i(1)|S_{i}^{(1)}=3) \sim \Norm(2.5,0.08)$. The regression parameters for Y-model are $\theta^{(0)}=(1,2)$ and $\theta^{(1)}=(1,1.2,-0.8,0.5)$.

\vspace{0.25cm}
\noindent {\em Scenario 4:} This scenario corresponds to Scenario 2 with closer atoms for the strata and different variances. In particular, the dissociative stratum has $(P_i(0)|S_{i}^{(0)}=1) = (P_i(1)|S_{i}^{(1)}=1) \sim \Norm(1.5,0.12)$, the associative positive stratum has $(P_i(0)|S_{i}^{(0)}=2)  \sim \Norm(2,0.1)$ and $(P_i(1)|S_{i}^{(1)}=3) \sim \Norm(2.5,0.08)$, and the associative negative stratum has $(P_i(0)|S_{i}^{(0)}=2)  \sim \Norm(2,0.1)$ and $(P_i(1)|S_{i}^{(1)}=1) \sim \Norm(1.5,0.12)$. The regression parameters for Y-model are $\theta^{(0)}=(1,2)$ and $\theta^{(1)}=(1,1.2,-0.8,0.5)$.

\vspace{0.25cm}
\noindent {\em Scenario 5:} We investigate the scenario with the three strata when the number of confounders increases, in particular the treatment variable and cluster allocation variables that depend on five confounders. The dissociative stratum has $(P_i(0)|S_{i}^{(0)}=1) = (P_i(1)|S_{i}^{(1)}=1) \sim \Norm(2,0.05)$, the associative positive stratum has $(P_i(0)|S_{i}^{(0)}=3)  \sim \Norm(3,0.05)$ and $(P_i(1)|S_{i}^{(1)}=4) \sim \Norm(4,0.05)$, and the associative negative stratum has $(P_i(0)|S_{i}^{(0)}=2)  \sim \Norm(2,0.05)$ and $(P_i(1)|S_{i}^{(1)}=1) \sim \Norm(1,0.05)$. The regression parameters for Y-model are $\theta^{(0)}=(1,2)$ and $\theta^{(1)}=(1,1.2,-1,0.5)$.

\vspace{0.25cm}

We choose the same hyperparameters for each setting such that the prior is non-informative and in common for all the settings. For the regression parameters in \eqref{eq:model4_p2} and for the parameters $\eta_l$ and $\sigma_l$ in \eqref{eq:model5_p2} we use the following conjugate priors
\begin{align*}
    &\beta^{(t)} \sim \Norm_{(p+1)(L-1)}(0,20\times I_{(p+1)(L-1)}),& \\
&\eta_l \sim \Norm(0,20), \mbox{ and } \sigma_l \sim \mbox{InvGamma}(2,0.5),&
\end{align*}
for $t\in \{0,1\}$, $l\in\{1,\dots,20\}$, and $p$ according with the covariates considered in different settings, and where $I_q$ is a diagonal matrix $q\times q$.

Table \ref{table: bias_P2} reports the median and interquartile range of the bias for the expected value of the posterior distribution of the sample average of $P_i(1)-P_i(0)$ and $Y_i(1)-Y_i(0)$, for $i \in \{1,\dots, n\}$. 

The results for the five scenarios demonstrate the strong ability of CASBAH to impute missing variables and accurately capture the true distribution across different data generating processes. The medians of the bias are close to zero and the interquartile range is reasonable given the simulated variability. In contrast, the two competing models highlight their limitations in recovering the same information. The algorithm from \citet{lu2023principal} does not estimate the distribution of the potential outcome for the post-treatment variable, while the model from \citet{schwartz2011bayesian} exhibits significant bias in four out of five scenarios and a higher interquartile range than CASBAH across all scenarios.
Furthermore, the comparison of $\E[Y_i(1)-Y_i(0)]$ reveals substantial bias in all five scenarios for both competing models \citet{lu2023principal} and \citet{schwartz2011bayesian}.

\begin{table}
	\caption{Bias comparison of the three methods based on different simulation scenarios. (IQR: interquartile range.)}
    \vspace{-0.7cm}
{	\footnotesize
\begin{tabular}{llrccccc}
\multicolumn{8}{c}{}\\
	%\hline
 &&& Scenario 1 & Scenario 2 & Scenario 3 & Scenario 4 & Scenario 5 \\ 
  %\hline
  \multicolumn{1}{c}{$\E\{P(1)-P(0)\}$} &&&&&&& \\
  \hline
  %\rowcolor{lightGray}
 %\cellcolor{white}
%Our proposed method & Median &&  -0.0164 & 0.0027 & -0.0318 & 0.0027 & 0.0011 \\ 
CASBAH & Median &&  -0.0142 &  0.0010 & -0.0032 & 0.0002 & 0.0019  \\ 
%\rowcolor{Gray2}
 %\cellcolor{white}
% & IQR && 0.0315 & 0.0254 & 0.0462 & 0.0325 & 0.0183 \\ 
  & IQR && 0.0235 & 0.0159 &  0.0285 & 0.0234 & 0.0193 \\ 
 %\rowcolor{lightGray}
 %\cellcolor{white}
Schwartz et al. (2011) &  Median && -0.0637 & 0.3388 & 0.1254 & 0.3280 & 0.3572 \\ 
%\rowcolor{Gray2}
 %\cellcolor{white}
 & IQR && 0.0989 & 0.0986 & 0.0956 & 0.0963 & 0.1216 \\ 
Lu et al. (2024) &  Median && -- & -- & -- & -- & -- \\ 
%\rowcolor{Gray2}
 %\cellcolor{white}
 & IQR && -- & -- & -- & -- & -- \\ 
   %\hline
\multicolumn{1}{c}{$\E\{Y(1)-Y(0)\}$} &&&&&&& \\
  \hline
  %\rowcolor{lightGray}
 %\cellcolor{white}
%Our proposed method & Median && 0.0024 & 0.0256 & 0.0067 & 0.0419 & 0.0344 \\ 
    CASBAH & Median && -0.0534 & -0.0001 & -0.0159 & -0.0027 &  0.0020 \\ 
%\rowcolor{Gray2}
 %\cellcolor{white}
 %& IQR && 0.2235 & 0.1550 & 0.4661 & 0.1357 & 0.1316 \\ 
  & IQR && 0.0699 & 0.1049 & 0.0885 & 0.0909 & 0.0720 \\ 
 %\rowcolor{lightGray}
 %\cellcolor{white}
Schwartz et al. (2011) &  Median && -1.9891 & -1.8265 & -1.4363 & -1.3115 & -1.8856 \\ 
%\rowcolor{Gray2}
 %\cellcolor{white}
 & IQR && 0.2546 & 0.2110 & 0.1904 & 0.1836 & 0.3727 \\ 
%\hline
Lu et al. (2024) &  Median && -1.3491 & 0.3070 & -0.5848 & 0.1393 & 0.0751 \\ 
%\rowcolor{Gray2}
 %\cellcolor{white}
 & IQR && 0.1092 & 0.1206 & 0.1009 & 0.0869 & 0.1078 \\
	\end{tabular}}
	\label{table: bias_P2}
	%\label{tab2}
\end{table}

To assess the accurate identification of the principal strata, we use the adjusted Rand index \citep{hubert1985comparing} and evaluate the estimation of the principal causal effects. The adjusted Rand index values for the five scenarios are reported in Table \ref{table: rand_p2}. For all scenarios, the index is close to $1$ and confirms that CASBAH can correctly identify the principal strata and, combined with the good missing data imputation, allows us to estimate the expected associative and dissociative effects.  Table \ref{table:PCE_sim_comp} shows that CASBAH correctly identifies the number of principal strata---two in Scenario 1 and 3, three in the others---while also providing accurate causal effect estimates. In contrast, neither the model by \citet{lu2023principal} nor that of \citet{schwartz2011bayesian}identify the principal strata according to our definition.% ---in the case of \cite{lu2023principal}--- or without further assumptions about the deinition of the principal casual effect ---in the case of \cite{schwartz2011bayesian}--- in opposition to our proposed model.

\begin{table}
	\caption{Adjusted rand index for the five simulated scenarios computed on the point estimated partitions obtained with our proposed model.}
        \vspace{-0.7cm}
{\footnotesize
\begin{tabular}{lrccccc}
\multicolumn{7}{c}{}\\
  %\hline
 && Scenario 1 & Scenario 2 & Scenario 3 & Scenario 4 & Scenario 5 \\ 
  \hline
  %\rowcolor{lightGray}
Mean && 0.9850 & 0.9906 & 0.9706 & 0.9717 & 0.9154 \\
%\rowcolor{Gray2}
Standard deviation && 0.0997 & 0.0597 & 0.1021 & 0.0892 & 0.1577 \\
   %\hline
\end{tabular}}
\label{table: rand_p2}
\end{table}

\begin{table}
\caption{True and estimated principal causal effects---mean and standard deviation in the brackets---from CASBAH model across scenarios. (NA: Not applicable. The stratum was not present in the scenario.)}
%\adjustbox{width=\textwidth}
    \vspace{-0.7cm}
{%
{\footnotesize
\begin{tabular}{lccccccccccc}
\multicolumn{11}{c}{}\\
%\hline
 & \multicolumn{2}{c}{Scenario 1} & \multicolumn{2}{c}{Scenario 2} & \multicolumn{2}{c}{Scenario 3} & \multicolumn{2}{c}{Scenario 4} & \multicolumn{2}{c}{Scenario 5} \\
 & True & Estimated & True & Estimated & True & Estimated & True & Estimated & True & Estimated \\
\hline
$\tau_{-}$ & N/A & -- & $-1.00$ & $-0.98$ (0.32) & N/A & -- & $1.50$ & $1.46$ (0.33) & $-1.00$ & $-0.79$ (0.68) \\
$\tau_0$ & $-0.5$ & $-0.50$ (0.24) & $4.00$ & $4.03$ (0.61) & $5.60$ & $4.98$ (0.52) & $4.00$ & $4.11$ (0.53) & $4.00$ & $4.57$ (1.50) \\
$\tau_{+}$ & $3$ & $2.91$ (0.11) & $9.00$ & $9.02$ (0.45) & $8.60$ & $9.36$ (0.87) & $6.50$ & $6.59$ (0.35) & $9.00$ & $8.91$ (0.44) \\
%\hline
\end{tabular}}%
}
\label{table:PCE_sim_comp}
\end{table}

%\begin{table}
%	\tbl{Estimation of the principal causal effects.}
%{	\begin{tabular}{lrcccc}
	%\hline
%  &&\multicolumn{2}{c}{Our proposed method}&\multicolumn{2}{c}{Lu et al. (2024)}\\
%   && mean & SD & mean & SD\\
%   \hline
% Scenario 1   &&&&&\\
% $\tau_{-}$ NA    & & -- & --&-1.99 & 0.17\\
%  $\tau_0 = 0.5$   &&0.11 & 0.56& -1.26 & 0.19\\
%    $\tau_{+} = 9$     &&8.79 & 1.25& 2.75 & 0.10\\
%Scenario 2    &&&&&\\
 %   $\tau_{-} = -1$     &&-0.98 & 0.32& -2.28 & 0.24\\
%$\tau_0 = 4$   &&4.03 & 0.61& -1.75 & 0.23\\
 %   $\tau_{+} = 9$     &&9.02 & 0.45& 3.20 & 0.07\\
%Scenario 3    &&&&&\\
 %$\tau_{-}$ NA     &&--&--& -1.99 & 0.12\\
 % $\tau_0 = 5.6$   &&4.98 & 0.52& -1.49 & 0.26\\
 %   $\tau_{+} = 8.6$     &&9.36 & 0.87& 2.13 & 0.11\\
%Scenario 4    &&&&&\\
 %   $\tau_{-} = 1.5$     &&1.46 & 0.33& -2.05 & 0.14\\
%$\tau_0 = 4$   &&4.11 & 0.53& -1.56 & 0.30\\
%    $\tau_{+} = 6.5$     &&6.59 & 0.35 & 2.20 & 0.11\\
%Scenario 5    &&&&&\\
%    $\tau_{-} = -1$     && -0.79 & 0.68& -2.98 & 0.31\\
%$\tau_0 = 4$   &&4.57 & 1.50& -1.66 & 0.18\\
%    $\tau_{+} = 9$    & &8.91 & 0.44& 2.43 & 0.08\\
%	\end{tabular}}
%	\label{table: PCE_sim_comp}
%	\begin{tabnote}
%	SD: standard deviation.
%	\end{tabnote}
%\end{table}

%% file: Sections/5_application.tex
\subsection{Data and Study Design}

In our application, we investigate the repercussions of the National Ambient Air Quality Standards (NAAQS) revision for air pollution in the U.S. This revision dictated a more stringent environmental policy for the concentration of \PMns. First, we evaluate the direct and indirect effects of the 2005 NAAQS revision on the mortality rate in the period 2010-2016, considering the variation of levels \PMns, under the principal stratification framework. Second, we take advantage of the flexible Bayesian non-parametric mixture model for the post-treatment variable (that is, \PMns) to understand how these effects can vary across principal strata. Third, we provide a characterization of these different identified groups.

To answer our research question, we have merged two datasets: (i) one dataset containing the information about the NAAQS designations, as well as \PM and the demographic and socioeconomic characteristics in the counties in the Eastern U.S., used in the \citet{zigler2018impact}'s analysis \cite[data can be found at][]{DVN/ZAYLFA_2017}, and (ii) one dataset containing the information on the age-adjusted mortality rate in these counties available by the Center for Disease Control and Prevention (CDC) of U.S. \citep{friede1993cdc}.

The initial dataset from \citet{DVN/ZAYLFA_2017} is made up of 482 counties in the Eastern U.S., where national monitoring networks have detected the concentration of \PMns. In 2005, the EPA designated as \textit{non-attainment} of the NAAQS these counties where the average concentration of \PM was above $15 \mu g/m^3$, or otherwise \textit{attainment} (note that the revision occurred in 1997, but became effective just in 2005 due to legal disputes). States containing counties designated as non-attainment were required to develop or revise State Implementation Plans (SIP) outlining how a non-attainment area will achieve standards with strategies to reduce ambient concentrations of \PMns. 

As noted in the previous literature, there is a large diversity in local actions in response to the non-attainment designation and in the enactment of subsequent SIPs \citep{greenstone2004did, zigler2018impact}. Due to this diversity, the direct analysis of the effect of the designation is very similar to an \textit{intention-to-treat} analysis. This highlights the importance of conducting a principal stratification analysis to compare the effects on health outcomes across different locations: specifically, it contrasts the effects in areas where pollution was effectively reduced by the non-attainment designation (associative negative effects) with those areas where pollution was not measurably affected (dissociative effect), or where it actually increased (associative positive effects).

The study design specifies a \textit{baseline period} from 2000 to 2005. During this time, each county, $i$, in the dataset is classified according to its attainment status under the EPA 2005 NAAQS revisions. We define a binary treatment variable, $T_i$, where $T_i=1$ indicates that the $i$-th county was designated as \textit{non-attainment} and therefore was required to develop or revise its State Implementation Plans. In contrast, $T_i=0$ indicates that the $i$-th county was designated as \textit{attainment} and did not have these requirements.

For this period, we also have the average ambient concentration of \PM (from the EPA monitoring locations within each county), as well as a number of counfounders such as census variables like: the percentage of Hispanic and black residents; the average household income; the percentage of females; the average house value; the proportion of residents in poverty; the proportion of residents with a high school diploma; the smoking rate; the population; the percentage of residents in urban area; the employment rate (the percentage of the workforce employed); the percentage of the move in the last 5 years. We also have meteorological variables such as the averages of daily temperatures and the relative average of humidity; the dew point (the temperature at which air becomes saturated with moisture, leading to the formation of dew or condensation).

Consistently with \citet{zigler2018impact}, we identify as the \textit{follow-up period} a period after 5 years from the implementation of the 2005 NAAQS revision. For this period, we have data on the levels of \PM (2010-2012) from \citet{DVN/ZAYLFA_2017}. The levels of \PM in the follow-up period are compared with the levels in the baseline period to establish the decrease / increase in air pollution following the implementation of NAAQS. This variable serves as the post-treatment variable $P_i$. Furthermore, we gather the publicly available age-adjusted mortality rate of all-cause mortality (2010-2016) from the website of the US Center for Disease Control and Prevention (CDC) \citep{friede1993cdc}. Mortality rates in the follow-up period are also contrasted with the rates in the baseline period to assess the impact of NAAQS on mortality rates accounting for baseline rates. This variable serves as our outcome variable $Y_i$.

The final dataset, obtained by merging the previously mentioned data sources and cleaning the data set to avoid missing data, comprises 384 counties, 270 of which were designed as attainment and 114 as non-attainment. In the Supplementary Material, Figure F.1 maps this final analysis data set. 

\subsection{Results}
\label{subsec:results}

We apply CASBAH to the 384 counties in the Eastern US described above, including all covariates, census and meteorological variables, in the weights of the post-treatment variable mixture, while for the outcome model we use the linear model in \eqref{eq:Y_model}. The model identifies the three strata: the dissociative stratum with 124 counties ($32\%$ of the total counties analyzed), the associative positive stratum with 46 counties ($12\%$), and the associative negative stratum with 214 counties ($56\%$). 

\begin{figure}%[!htb]
\centering
\includegraphics[height=1.8in]{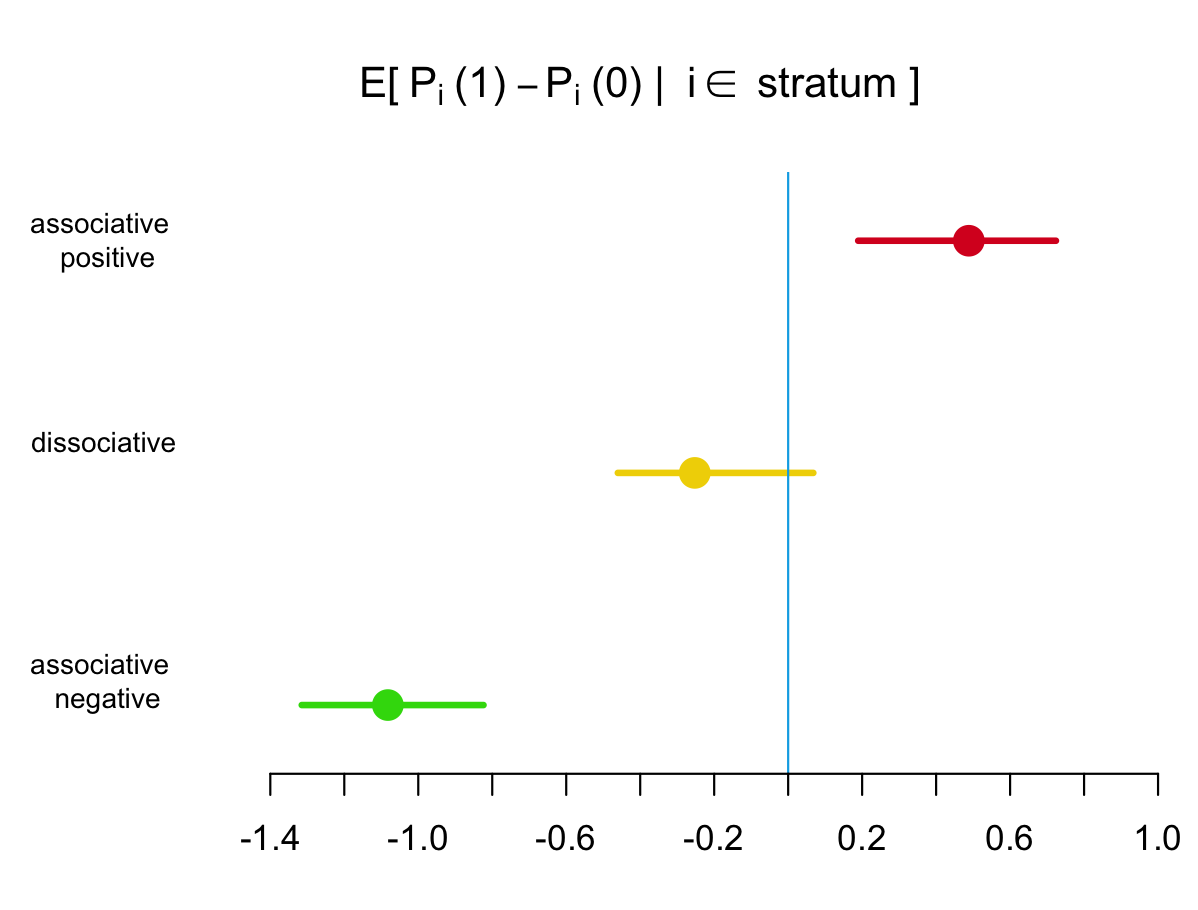} \includegraphics[height=1.8in]{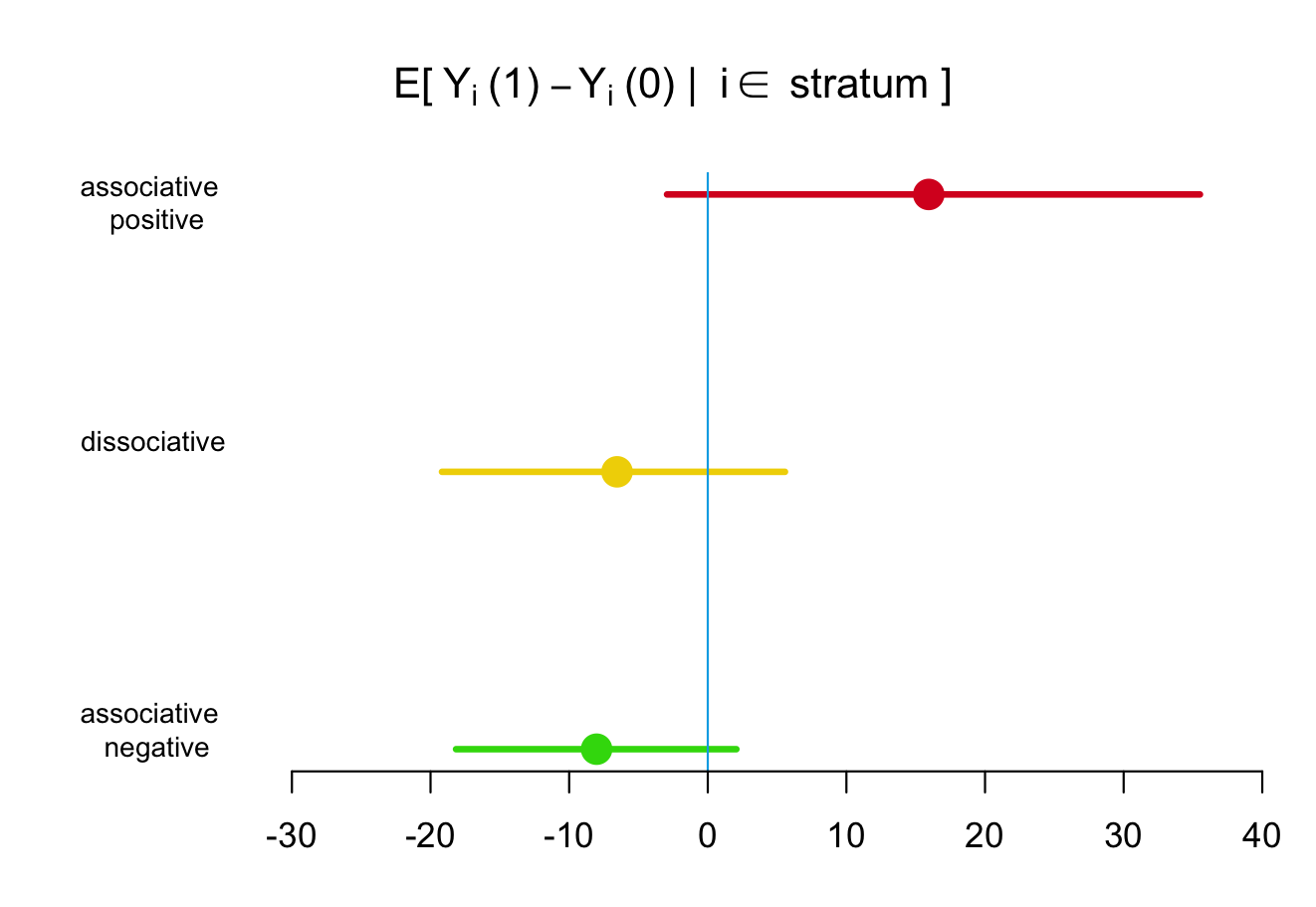}
\caption{Posterior medians and $90\%$ credible intervals of the three identified strata for (left) the conditional average of the difference of the post-treatment variables, and (right) the expected associative/dissociative effects. The x-axes indicate (left) the \PM variation in $\mu g/m^3$ and (right) the mortality variation in $\percentomila$. The light-blue vertical lines show the value zero, identifying the null effects.}
\label{fig: strata}
\end{figure}

According to the definition of the three strata and as visualized in the image on the left in Figure \ref{fig: strata}, the dissociative stratum, identified with the color yellow, is composed of counties where the NAAQS revision does not substantially affect the level of \PMns, in fact, the expected value of $\E\{P_i(1)-P_i(0) \mid S_{i}^{(1)} = S_i^{(0)}\}$ for the counties allocated to this strata has a median close to zero and the $90\%$ credible intervals of $-0.46 \mu g/m^3$ and $0.07  \mu g/m^3$. The associative negative stratum, identified with the color green, is made up of counties where the implementation of environmental plans significantly decreases the levels of \PMns. Specifically, in these counties, the NAAQS revision reduced by $-1.09 \mu g/m^3$ the median levels \PMns, with $90\%$ credible intervals of of $-1.32 \mu g/m^3$ and $-0.82  \mu g/m^3$. The associative positive stratum, identified with the color red, is made up of counties where the revision of NAAQS was associated with increases in \PM levels by $0.50 \mu g/m^3$ in the median and credible intervals equal to $0.19 \mu g/m^3$ and $0.72  \mu g/m^3$, respectively.

The corresponding distributions of the expected dissociative/associative effects are reported in the right image in Figure \ref{fig: strata} and show the effect of the attainment or non-attainment designations on the mortality rate conditional to the three strata, i.e., conditional to the heterogeneity in the causal effects of the NAAQS revision on the level of pollution. The associative negative effect, in green color on the right image of Figure \ref{fig: strata}, assumes negative values, indicating that the implementation of environmental plans, in the counties where the regulations affect the level of \PMns, reduce it, also reduces the median age-adjusted mortality rate. Specifically, the mortality rate decreases by $8.12\percentomila$ in the median. %and with credible intervals of $[-20\percentomila,-11.2\percentomila]$. 
The dissociative effect (in yellow) shows a decrease of $8.16\percentomila$ in median of the mortality rate when environmental plans are applied, while the associative positive effects decrease in median by $17.87\percentomila$. %The credible intervals of the principal effects have credible intervals that include the zero value. This means that in counties where the NAAQS revision does not affect the level \PM or where this level increases, the mortality rate is also not affected. 

%Considering the high variability of the variation of the mortality rate between baseline and follow-up period, with the interquartile range of $65\percentomila$, the mean values of the expected dissociative effect and the expected associative positive effect can be considered close to zero, i.e., the null effect, while the median value of the expected associative negative effect can be considered significantly different from zero.

\begin{figure}%[!htb]
\begin{center}
\includegraphics[trim={1cm 1.5cm 1cm 1.5cm}, width=2.5in]{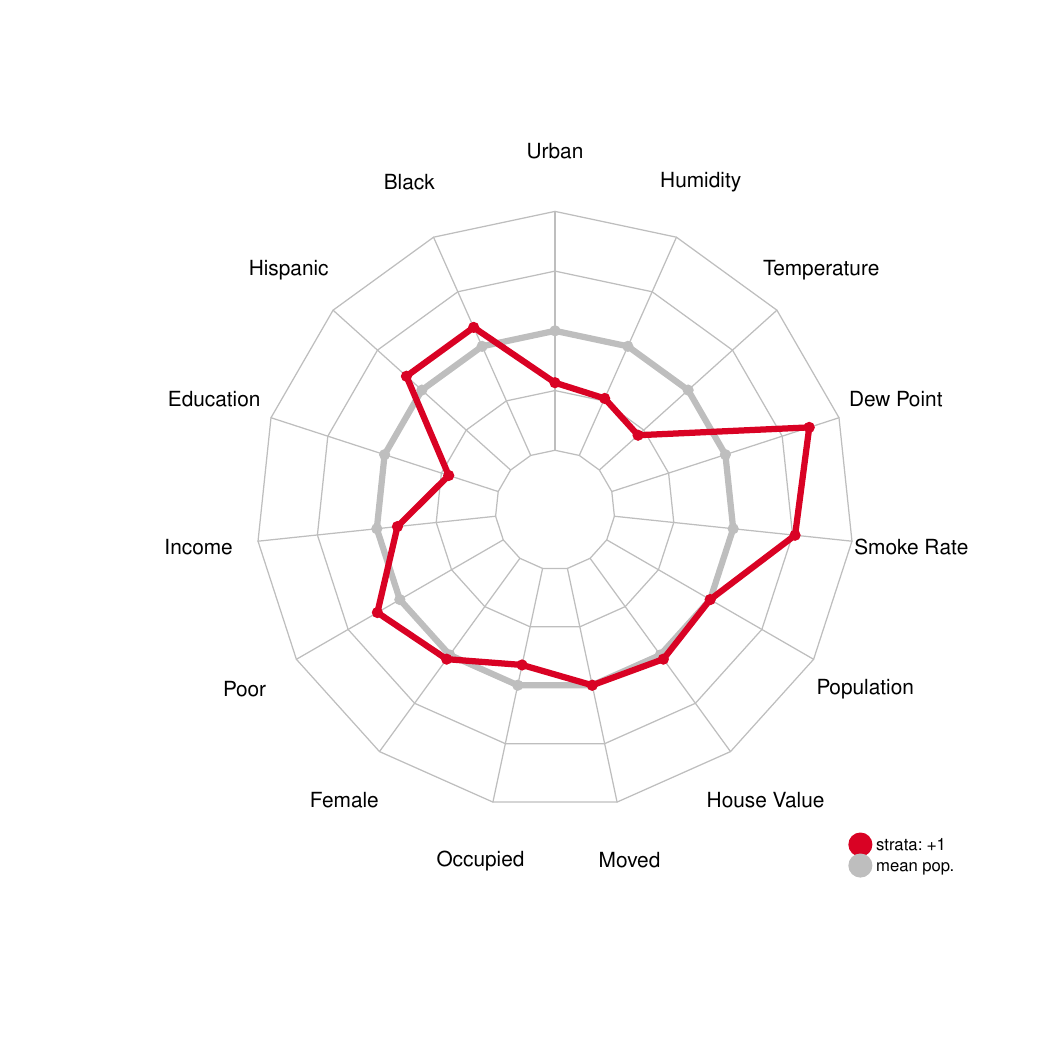}\;\includegraphics[trim={1cm 1.5cm 1cm 1.5cm},width=2.5in]{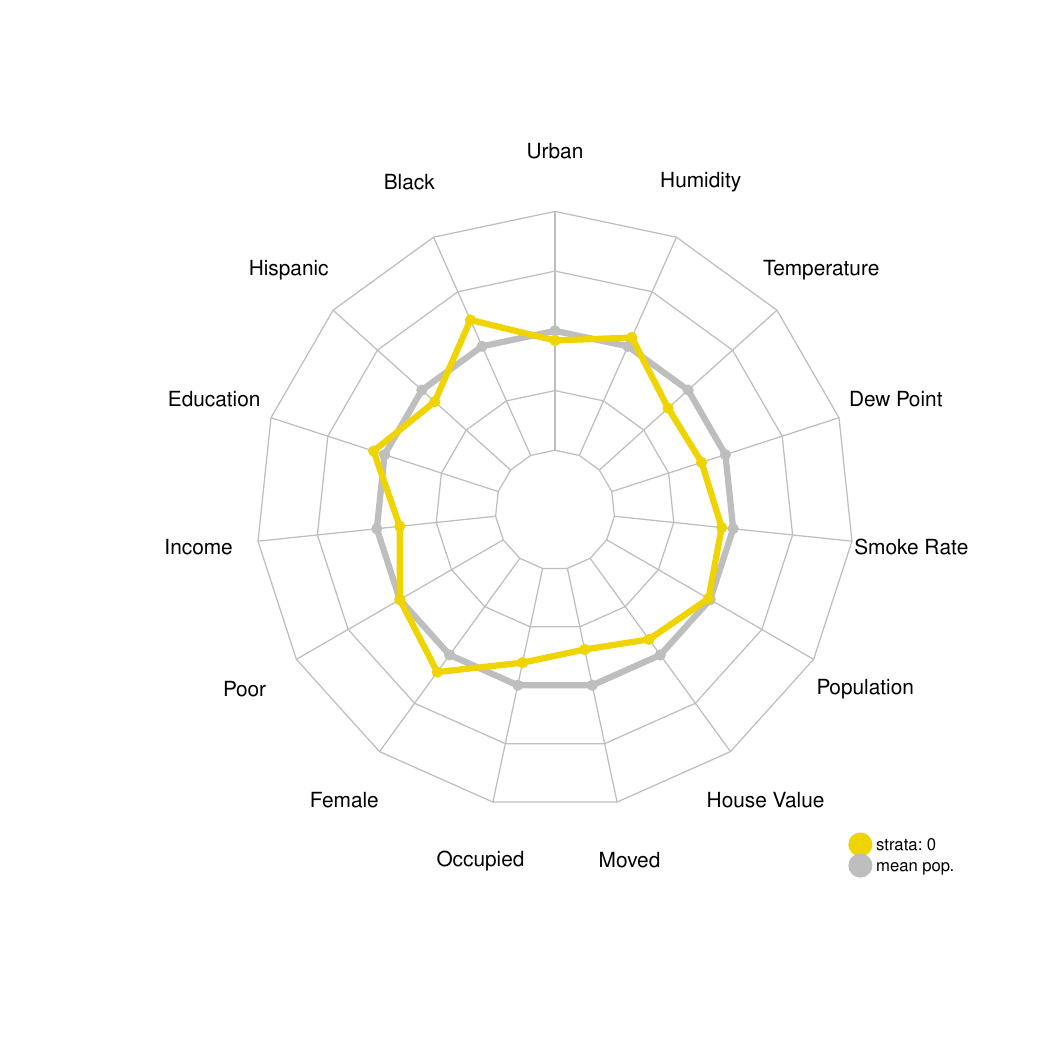}\\ \includegraphics[trim={1cm 1.5cm 1cm 1.5cm},width=2.5in]{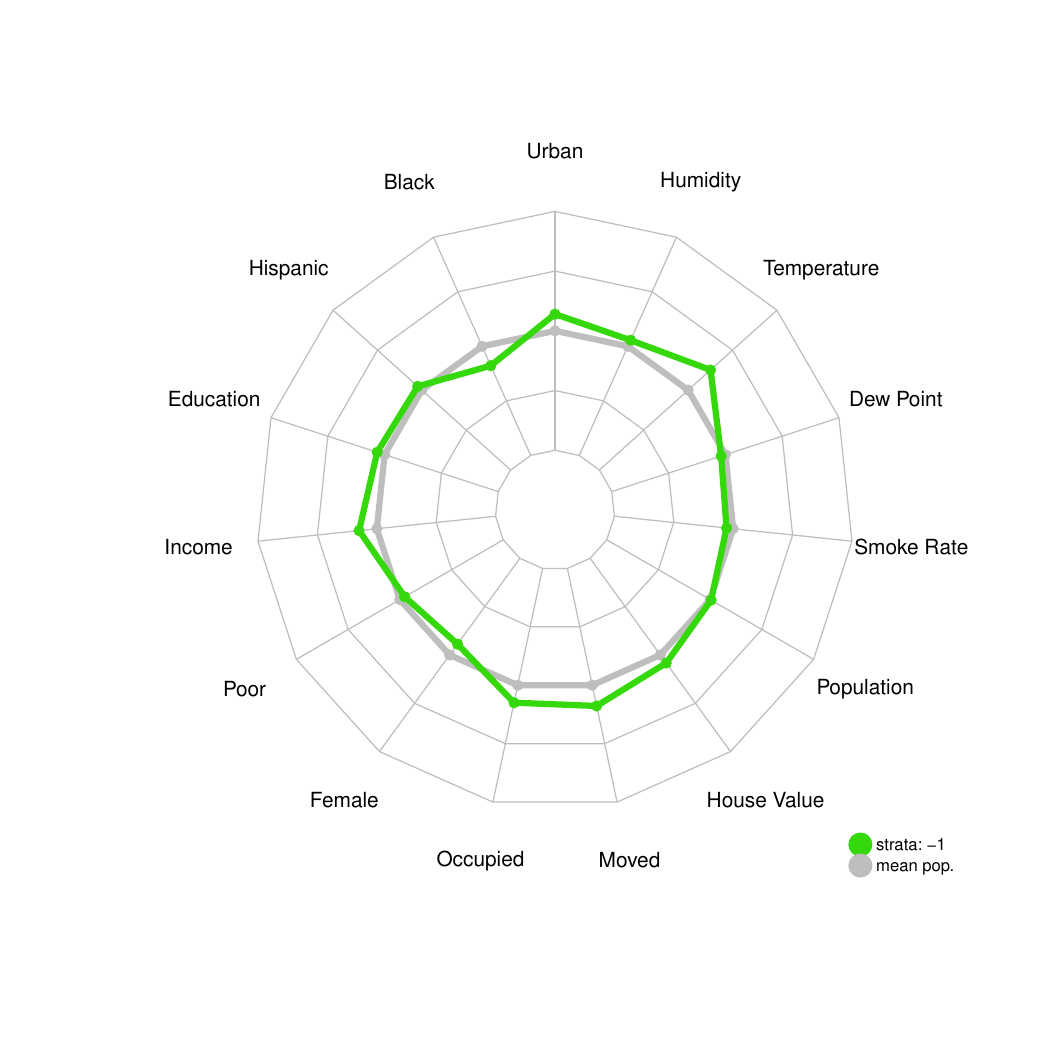}
\end{center}
\caption{Representation of the characteristics of the identified strata. Each spider plot reports in the colored area the strata-specific characteristics (the mean of the analyzed covariates) and in the gray area the collective characteristics (the mean of the covariates among all the analyzed counties in the Eastern U.S.). We can consider the gray area as the benchmark to understand how the characteristics of each stratum differ from the collective characteristics of the analyzed population. \label{fig:covariates_P2}}
\end{figure}

In addition, it is our interest to characterize the three different strata. Figure\ref{fig:covariates_P2} visualizes the average of observed covariates within each stratum, reported as colored lines, and compares them with the average of covariates between the full 384 counties in gray. The associative positive stratum and the dissociative stratum are composed of rural counties with a higher percentage of women and Black communities, where the population has lower income relative to the mean of the overall counties and with a small employment rate. In addition, the positive associative stratum (in red) is also characterized by a lower education rate, higher Hispanic community and a higher percentage of smokers.

In contrast, the counties in the associative negative stratum (in green) are mainly composed by urban areas with a high population density, high levels of education, higher income and house values, higher rate of move, and more men. The meteorological variables, such as the averages of daily temperatures, the relative average of humidity, and the dew point, also appear to play an important role in the positive associative stratum---that is, in the characterization of the different effects of the implementation of environmental plans on the level of \PMns. We refer to the Supplementary Materials for more details about the probability of each county to belong in each of the three strata.

%In addition, our proposed approach allows us to quantify the uncertainty of strata allocation. In fact, for each county, we know the probability of being allocated in each of the three strata, in addition to the estimation of its allocation. In addition, we can visualize this information on the US map. Specifically, the first three maps in Figure \ref{fig:map_strata} visualize the probability that each county is assigned to the three different strata. As already underlined in Figure \ref{fig:covariates_P2}, counties with a higher probability of being allocated in the associative negative stratum and the dissociative stratum are far from the largest cities, different from the associative positive stratum. In addition, western countries seem to have a small probability of being assigned to the associative negative stratum. The fourth map---bottom right in Figure \ref{fig:map_strata}---reports the estimation of the partition point of the strata, a partition that is used to estimate the principal causal effects.

%% file: Sections/6_conclusion.tex
In this paper, we proposed the Confounders-Aware SHared Atoms Bayesian Mixture Model (CASBAH) to address critical challenges within the principal stratification framework. Our approach leverages the dependent Dirichlet process to define a highly flexible and adaptable model for the potential post-treatment variables. An innovative feature of CASBAH is that we allow the distribution of potential post-treatment variables to share information through the treatment levels via the shared-atoms of the Bayesian nonparametric prior. This enables a more accurate identification of the principal strata while relaxing the standard assumptions of the causal framework of the principal stratification. 

CASBAH data-adaptively identifies the principal strata and imputes missing post-treatment variables and outcomes without the need to set predefined (and subjective) thresholds on the continuous post-treatment values to identify the principal strata. Moreover, the proposed model allows for the quantification of the uncertainty in membership in the principal strata. Theorem 1 provides identification for the proposed causal estimands, while Theorem 2 illustrates how the probability of a unit belonging to a dissociative stratum is nonzero a priori and depends on its observed covariates.

The performance of the proposed model is evaluated through an extensive set of Monte Carlo simulations where the performance of CASBAH is compared and contrasted with the performance of the proposed models by \citet{schwartz2011bayesian} and \citet{lu2023principal}. The results demonstrate that CASBAH consistently outperforms these alternatives, achieving lower bias, better principal strata identification, and accurate estimation of principal causal effects across various scenarios.

The proposed model is used to assess the effectiveness of a previous revision of NAAQS in reducing levels of \PM and, in turn, reducing the mortality rate in treated counties.  Specifically, we find significant effects on the reduction of mortality rate in areas where pollution was effectively reduced by the non-attainment designation (associative negative effects). In contrast, those areas where pollution was not measurably affected (dissociative effect) or where it actually increased (associative positive effects) did not show any significant effect on the mortality rate.

Future research could build upon our methodology by adopting more flexible approaches to model the primary outcome. While our current model uses a linear specification for the outcome, future work could explore non-linear and more sophisticated modeling techniques. In this study, we focus on innovations in modeling the post-treatment variable and intentionally keep the outcome model simpler to avoid unnecessary complexity. However, exploring more complex options for outcome models remains an opportunity for further research. Furthermore, our methodology could be adapted to accommodate continuous treatments, similar to the approach suggested by \citet{antonelli2023principal}, or extended to handle the time-to-event setting as in \citet{xu2022bayesian}. We also leave the exploration of these extensions to future studies.

%% file: Appendix/appendix_sun_dist.tex
%\subsection{Conjugate Priors for Multinomial Probit Models}
\label{subsec:sun}

The choice to assume a $\sun$ distribution \citep{arellano2006unification} as prior for regression parameters in the probit regression, Eq.~(6), is due to it is the conjugate prior for probit regression and consequently, it allows us (i) to obtain an efficient step in the Gibbs sampler and (ii) to avoid data augmentation that has usually same drawbacks. 

The $\sun$ density distribution for $\beta^{(t)}$ in Eq.~(6) is defined, for $q=(p+1)(L-1)$, as
\begin{equation}
    \P(\beta^{(t)})=\phi_{q}(\beta^{(t)} - \xi;\Omega)\frac{\Phi_h(\gamma+\Delta^T \Bar{\Omega}^{-1}\omega^{-1}(\beta^{(t)}-\xi);\Gamma-\Delta^T \Bar{\Omega}^{-1}\Delta)}{\Phi_h(\gamma;\Gamma)},
    \label{eq:pr_beta_sun}
\end{equation}
where $\phi_{q}(\beta^{(t)} - \xi;\Omega)$ is a $q$-variate Gaussian distribution with $\xi$ vector of location parameters and $\Omega$ the covariance matrix, such that $\Omega=\omega \Bar{\Omega} \omega$ where $\Bar{\Omega}$ is the correlation matrix and $\omega=(\Omega\bigodot \mathbf{1}_q)^{1/2}$ where $\bigodot$ is the element-wise Hadamard product.. The second part of the formula introduces a skewness mechanism, driven by the cumulative distribution function, computed at $\gamma+\Delta^T \Bar{\Omega}^{-1}\omega^{-1}(\beta^{(t)}-\xi) \in \mathbb{R}^h$ of an $h$-variate Gaussian with mean vector $0$ and $h\times h$ covariance matrix $\Gamma-\Delta^T \Bar{\Omega}^{-1}\Delta$. The quantity $\Phi_h(\gamma;\Gamma)$ is the normalizing constant, which coincides with the cumulative distribution function, evaluated at $\gamma \in \mathbb{R}^h$, of
an $h$-variate Gaussian with mean vector $0$ and $h\times h$ covariance matrix $\Gamma$.

The amount of skewness in the prior is mainly controlled by the $q\times h$ matrix $\Delta$, and when all the entries in $\Delta$ are $0$, the prior for $\beta_t$ coincides with the density of a $q$-variate Gaussian distribution with $\xi$ vector of location parameters and $\Omega$ the covariance matrix \citep{fasano2022class}.

\citet{arellano2006unification}  show that if $\beta^{(t)} \sim \sun_{q,h}(\xi,\Omega,\Delta,\gamma,\Gamma)$ then 
\begin{align*}
    \beta^{(t)} &\stackrel{d}{=} \xi +\omega(B_0^{(t)}+\Delta\Gamma^{-1}B_1^{(t)}),\\
    B_0^{(t)} &\sim \mathcal{N}_q(0,\Bar{\Omega}-\Delta\Gamma^{-1}\Delta^T),\\
    B_1^{(t)} &\sim TN_h(-\gamma;0,\Gamma),
\end{align*}
where $TN_h(-\gamma;0,\Gamma)$ denotes an $h$-variate Gaussian with zero mean, covariance matrix $\Gamma$ and truncation below $-\gamma$. A simple mechanism that helps in the simulation of $\sun$ variables.

The multinomial probit distribution for the weights $\pi^{(t)}=\{\pi_{l}^{(t)}\}_{l=1}^L$ can be rewritten as
\begin{equation*}
    \P(S_{i}^{(t)}=l|\beta^{(t)},X_i)=\Phi(x_i^T\beta_{l}^{(t)})\prod_{k=1}^{l-1}[1-\Phi(x_i^T\beta_{k}^{(t)})]=\prod_{k=1}^{l}\Phi\left((2\Bar{s}_{ik}^{(t)}-1)x_i^T\beta_{k}^{(t)}\right)=\Phi_l(x_i^T\beta^{(t)};I_l)
\end{equation*}
for $t=\{0,1\}$ and $l=1,\dots,L-1$, and where $x_i=(1,x_{i1},\dots,x_{ip}^T)$ is the vector of the $p$ covariates and intercept for the unit $i$, $\Bar{s}_{i}^{(t)}=(0_{S_{i}^{(t)}-1}^T,1)^T$ if $S_{i}(t) \leq L-1$ and $\Bar{s}_{i}^{(t)}=0_{L-1}$ if $S_{i}^{(t)} = L$, and $I_{l}$ refers to the $l\times l$ identity matrix.

Consequently, the probability over the observation $i=1,\dots,n_t$ for $t=\{0,1\}$ is 
\begin{equation}
    \P(S^{(t)}|\beta^{(t)},X)=\prod_{i=1}^{n^{(t)}} \P(S^{(t)}|\beta^{(t)},X_i) = \Phi_{\Bar{n}_t}(\Bar{X}^{(t)}\beta^{(t)},\mathbf{I}_{\Bar{n}^{(t)}})
    \label{eq:beta_likelihhod}
\end{equation}
 where $\Bar{n}^{(t)}=n_{1}^{(t)}+\dots+n_{n}^{(t)}$ with $n_{i}^{(t)}=\min(S_{i}^{(t)},L-1)$, $\Bar{X}^{(t)}$ is a $\Bar{n}^{(t)}\times[(p+1)(L-1)]$ matrix with row blocks $\Bar{X}^{(t)}_{[i]}=X_i^{(t)}$ and $X_i^{(t)}=(diag(2\Bar{s}_{i}^{(t)}-1)\bigotimes x_i^T,0_{(n_{i}^{(t)}\times[(p+1)(L-1-n_{i}^{(t)})])})$.

 Considering with the prior \eqref{eq:pr_beta_sun} and the likelihood \eqref{eq:beta_likelihhod}, the posterior distribution for $\beta^{(t)}$ is
\begin{equation}
    \P(\beta^{(t)}|S^{(t)},X)=\phi_{q}(\beta^{(t)} -\xi;\Omega)\frac{\Phi_{h+{\Bar{n}_t}}(\gamma_{pst}+\Delta_{pst}^T \Bar{\Omega}^{-1}\omega^{-1}(\beta^{(t)}-\xi);\Gamma_{pst}-\Delta_{pst}^T \Bar{\Omega}^{-1}\Delta_{pst})}{\Phi_{h+{\Bar{n}_t}}(\gamma_{pst};\Gamma_{pst})},
    \label{eq:post_beta}
\end{equation}
where $\Delta_{pst}=(\Delta,\Bar{\Omega}\omega(\Bar{X}^{(t)})^Td^{-1})$, $\gamma_{pst}=(\gamma^T,\xi^T(\Bar{X}^{(t)})^Td^{-1})$, $\Gamma_{pst}$ is an $(h+{\Bar{n}_t})\times(h+{\Bar{n}_t})$ covariance matrix with blocks $\Gamma_{pst[11]}=\Gamma$, $\Gamma_{pst[22]}=d^{-1}(\Bar{X}^{(t)}\Omega(\Bar{X}^{(t)})^T+\mathbf{I}_{\Bar{n}^{(t)}})d^{-1}$, and $\Gamma_{pst[12]}=\Gamma_{pst[21]}=d^{-1}\Bar{X}^{(t)}\omega\Delta$, where $d=[(\Bar{X}^{(t)}\Omega(\Bar{X}^{(t)})^T+\mathbf{I}_{\Bar{n}^{(t)}})\bigodot \mathbf{I}_{\Bar{n}^{(t)}}]^{1/2}$.

In the particular case in which the prior for $\beta^{(t)}$ is a multivariate Gaussian distribution, i.e. $h=0$, then the posterior is still the $\sun$ distribution in Eq.~\eqref{eq:post_beta} with $\Delta_{pst}=\Bar{\Omega}\omega(\Bar{X}^{(t)})^Td
^{-1}$, $\gamma_{pst}=d^{-1}\Bar{X}^{(t)}\xi$, and $\Gamma_{pst}=d^{-1}(\Bar{X}^{(t)}\Omega(\Bar{X}^{(t)})^T+\mathbf{I}_{\Bar{n}^{(t)}})d^{-1}$.

Moreover, a reasonable assumption for $\beta^{(t)}$ prior is the independence among the $q$ elements, such that $\Omega =\omega^2 \cdot\mathbf{I}_q$, i.e. the correlation matrix $\Bar{\Omega}=\mathbf{I}_q$. Therefore, following again the  \citet{arellano2006unification}'s results, the posterior distribution 
 of  $\beta^{(t)}$ can be drawn from 
\begin{align*}
    \beta^{(t)} &\stackrel{d}{=} \xi +\omega(B_{0,pst}^{(t)}+\Delta_{pst}\Gamma_{pst}^{-1}B_{1,pst}^{(t)}),\\
    B_{0,pst}^{(t)} &\sim \mathcal{N}_q(0,\mathbf{I}_q-\Delta_{pst}\Gamma_{pst}^{-1}\Delta_{pst}^T),\\
    B_{1,pst}^{(t)} &\sim TN_{h+{\Bar{n}^{(t)}}}(-\gamma_{pst};0,\Gamma_{pst}),
\end{align*}
with $\Delta_{pst}=\omega(\Bar{X}^{(t)})^Td
^{-1}$, $\gamma_{pst}=d^{-1}\Bar{X}^{(t)}\xi$, and $\Gamma_{pst}=d^{-1}(\omega^2\Bar{X}^{(t)}(\Bar{X}^{(t)})^T+\mathbf{I}_{\Bar{n}^{(t)}})d^{-1}$  where $d=[(\omega^2\Bar{X}^{(t)}(\Bar{X}^{(t)})^T+\mathbf{I}_{\Bar{n}^{(t)}})\bigodot \mathbf{I}_{\Bar{n}^{(t)}}]^{1/2}$.

%% file: Appendix/proof_diss_prob.tex
As defined in Section 3.1, a unit $i$ is allocated in the dissociative stratum when the two latent cluster allocation variables have same value, i.e., $S_i^{(0)}=S_i^{(1)}$. Given the prior distribution $G^{(0)}_{x_i},G^{(1)}_{x_i}$, the probability of unit $i$ to be allocated in the dissociative stratum is the following:

\begin{align}
    \Pr(S_i^{(0)}=S_i^{(1)}) & = \E\big\{\Pr(S_i^{(0)}=S_i^{(1)} \mid G^{(0)}_{x_i},G^{(1)}_{x_i}) \big\} \notag \\
    & = \E\bigg\{ \sum_{l=1}^L \pi_l^{(0)}(x_i) \pi_l^{(1)}(x_i) \bigg\}\notag \\
    & = \sum_{l=1}^L \E\big\{ \pi_l^{(0)}(x_i) \pi_l^{(1)}(x_i)\big\}\notag \\
    & = \sum_{l=1}^L \E\big\{ \pi_l(x_i)^2\big\},
    \label{eq:prop_1}
\end{align}
where the first and third equalities invokes the properties of expectation and the second equality invokes the definition of the latent variables $S_i^{(t)}$, for $t=\{0,1\}$, and the independence of the sequence of the random weights $\{\pi^{(0)}(x_i)\}_{l \geq 1}$ and $\{\pi^{(1)}(x_i)\}_{l \geq 1}$. Moreover the two sequence of the random weights are equivalent in distribution such that we can write $\pi_l^{(0)}(x) \stackrel{d}{=} \pi_l^{(1)}(x) \stackrel{d}{=} \pi_l(x)$, notation used in the fourth equality. 

By the definition of the probit stick-breaking, the expected value of the $l$-th weight squared is the follows
\begin{align*}
    \E\big\{ \pi_l(x_i)^2\big] & = \E\bigg[ \Phi\big(\alpha_l(x_i)\big)^2 \prod_{r < l} \bigg\{1-\Phi\big(\alpha_r(x_i)\big)\bigg\}^2 \bigg] \\
    & = \E\bigg[ \Phi\big(\alpha_l(x_i)\big)^2 \prod_{r < l} \bigg\{1-\Phi\big(\alpha_r(x_i))^2-2\Phi\big(\alpha_r(x_i)\big\}\bigg) \bigg] \\
    & = \E\big[ \Phi\big(\alpha_l(x_i)\big)^2\big] \prod_{r < l} \bigg\{1+\E\big[\Phi\big(\alpha_r(x_i))^2\big]-2\E\big[\Phi\big(\alpha_r(x_i)\big)\big]\bigg\} \bigg]
\end{align*}
For simplify the notation we will indicate $\rho_1(x_i)=\E\big\{\Phi\big(\alpha_l(x_i)\big)\big\}$ and $\rho_2(x_i)=\E\big\{\Phi\big(\alpha_l(x_i)\big)^2\big\}$, for each $l \in \{1, \dots, L\}$. Then \eqref{eq:prop_1} is rewritten as

\begin{align*}
    \Pr(S_i^{(0)}=S_i^{(1)}) & = \sum_{l=1}^L \E\big\{ \pi_l(x_i)^2\big\} \\
    & = \sum_{l=1}^L \bigg[\rho_2(x_i) \prod_{r < l} \big\{1+\rho_2(x_i)-2\rho_1(x_i)\big\}\bigg] \\
    & = \sum_{l=1}^L \bigg[\rho_2(x_i) \{1+\rho_2(x_i)-2\rho_1(x_i)\}^{l-1}\bigg] \\
    & = \frac{\rho_2(x_i)[\{1+\rho_2(x_i)-2\rho_1(x_i)\}^L-1] }{\rho_2(x_i)-2\rho_1(x_i)},
\end{align*}
where the third and fourth equalities are following by properties of mathematical series, and it is reported as Theorem 2. 

%% file: Appendix/appendix_Gibbs.tex
\label{subsec:gibbs_P2}

\citet{rodriguez2011nonparametric} proves that the finite truncation of the dependent Probit Stick-Breaking process is a good approximation; therefore, we can rewrite the model (3) as a finite mixture to $L<\infty$ components with $L$ a reasonable conservative upper bound. \citet{rodriguez2011nonparametric}'s proof is a key point that allows us to provide a simpler algorithm without losing the robustness of the model. 

In this section, we describe the Gibbs sampling algorithm for model fitting that allows us to draw from the posterior distribution. 
Following the steps in the algorithm~\ref{alg:common}, in each iteration $r=1,\dots, R$, we use the observed data $(y,p,t,x)$ to update the parameters and the augmented variables and impute the missing post-treatment variable $P^{mis}$ and missing outcome $Y^{mis}$.

The Gibbs sampling algorithm is divided into three parts: the estimation of the shared atoms mixture model for the post-treatment variables (divided in the estimation of cluster allocation, cluster-specific parameters, and confounder-dependent weights), the imputation of the missing post-treatment variables, and the estimation of the outcome model. 

As already discussed, the outcome model is not our main concern, therefore we assume a linear model. In particular, in the following Gibbs sampler, we consider the outcome model used in the simulation study, i.e., Eq. (10), where only the potential post-treatment variables are included. This choice is driven by the purpose of focusing attention on the essential definition of the relation between the post-treatment variable and outcome, which is crucial to impute the missing post-treatment variable $P^{mis}$. 
However, the algorithm can be easily modified to include the covariates $X$ in the linear regression or to consider a more complex and flexible model. 

\vspace{0.25cm}
\noindent {\em Cluster Allocation.}
The latent variables $S_{i}^{(t)}$ identifies the cluster allocation for each units $i \in \{1,\dots,n\}$ at the treatment level $t$. Its posterior distribution is a multinomial distribution where
\begin{equation*}
    \Pr\{S_{i}^{(t)}=l \} \propto \pi_{l}^{(t)}(x_i)\Norm(p_i;\eta_{l},\sigma_{l}^{2}),
\end{equation*}
for $i=1,\dots,n$ and $l=1,\dots,L$, with $\pi^{(t)}_{l}$ defined as:
\begin{equation*}
    \pi_{l}^{(t)}(x_i)=\Phi(\alpha_{l}^{(t)}(x_i)) \prod_{r<l} \{1-\Phi(\alpha_{r}^{(t)}(x_i))\},
\end{equation*} 
for $l=1,\dots,L-1$ and with $\Phi(\alpha_{L}^{(t)}(x_i))=1$.

\vspace{0.25cm}
\noindent {\em Cluster Specific Parameters.} 
Thanks to the latent variables $\{S_{i}^{(0)},S_{i}^{(1)}\}$, that cluster the units by the value of their outcome, we know for each cluster $l \in \{1,\dots, L\}$, the allocated units and we can update the values of the parameters from their posterior distributions:
\begin{align*}
    \eta_{l} & \sim \Norm\left(V_l^{-1}\times \left(\frac{ \sum_{\{i: S_{i}^{(0)}=l\}}p_{i}(0)+ \sum_{\{i: S_{i}^{(1)}=l\}}p_{i}(1)}{\sigma_{l}^{2}} +\frac{\mu_\eta}{\sigma^2_{\eta}} \right),V_l^{-1}\right);\\
    \sigma_{l}^{2} & \sim \mbox{InvGamma}\left(\gamma_1+ \frac{n_{l}}{2},\gamma_2+\frac{\sum_{\{i: S_{i}^{(0)}=l\}} (p_{i}(0)-\eta_{l})^2+\sum_{\{i: S_{i}^{(1)}=l\}} (p_{i}(1)-\eta_{l})^2}{2}\right);
\end{align*}
for $l=1,\dots,L$ and where $V_l=n_{l}/\sigma_{l}^{2} +1/\sigma^2_{\eta}$ and $n_{l}$ is the number of units allocated in the $l$-th cluster.

\vspace{0.25cm}
\noindent {\em Confounder-Dependent Weights.}
The $\{\beta_{ql}^{(t)}\}_{q=0}^p =(\beta_{0l}^{(t)},\beta_{l}^{(t)})$, for  $l=1,\dots,\max(S_{i}^{(t)},L-1)$, are updated for the posterior distribution:
\begin{align*}
    \beta^{(t)} &\stackrel{d}{=} \xi +\omega(B_{0,pst}^{(t)}+\Delta_{pst}\Gamma_{pst}^{-1}B_{1,pst}^{(t)}),\\
    B_{0,pst}^{(t)} &\sim \mathcal{N}_q(0,I_q-\Delta_{pst}\Gamma_{pst}^{-1}\Delta_{pst}^T),\\
    B_{1,pst}^{(t)} &\sim TN_{h+{\Bar{n}^{(t)}}}(-\gamma_{pst};0,\Gamma_{pst}),
\end{align*}
with $\Delta_{pst}=\omega(\Bar{X}^{(t)})^Td
^{-1}$, $\gamma_{pst}=d^{-1}\Bar{X}^{(t)}\xi$, and $\Gamma_{pst}=d^{-1}(\omega^2\Bar{X}^{(t)}(\Bar{X}^{(t)})^T+I_{\Bar{n}^{(t)}})d^{-1}$  where $d=[(\omega^2\Bar{X}^{(t)}(\Bar{X}^{(t)})^T+I_{\Bar{n}^{(t)}})\bigodot I_{\Bar{n}^{(t)}}]^{1/2}$. More details for $\Bar{X}^{(t)}$ and $\omega$ definitions in Section \ref{subsec:sun}.

\vspace{0.25cm}
\noindent {\em Imputation Missing Post-Treatment Variables.} For each unit $i \in \{1, \dots, n\}$, we impute the missing post-treatment variable $P_i^{mis}$. Firstly, drawing the relative cluster-allocation variable $S_i^{(1-t)}$, where $t$ is the observed treatment of the unit $i$, from a  multinomial distribution with
\begin{equation*}
    \mathbb{P}\{S_{i}(1-t)=l \} \propto \pi_{l}^{(1-t)}(x_i),
\end{equation*}
for $l=1,\dots,L$. Where $\pi_{l}^{(1-t)}(x_i)=\Phi(\alpha_{l}^{(1-t)}(x_i)) \prod_{r<l} (1-\Phi(\alpha_{r}^{(1-t)}(x_i)))$, for $l=1,\dots,L-1$ and with $\Phi(\alpha_{L}^{(1-t)}(x_i))=1$. 

Successively, drawing the missing post-treatment variable $P_i^{mis}$, conditioned to the allocation to the cluster $l$ and the observed outcome variables $Y_i(t)$. For each $i$ such that the observed treatment level is $T=1$, $P_i(1-t)$ is drown from 
\begin{equation*}
    \{P_i^{mis}| S_i(1-t)=l, \eta,\sigma^2, P_i, Y_i\} \sim \mathcal{N}\left(v^{-1} \left(\frac{\eta_l}{\sigma^2_l}+ \frac{m_1}{v_1} \right),v^{-1} \right);
\end{equation*}
where 
\begin{equation*}
    v = \frac{1}{\sigma^2_l}+\frac{1}{v_1}, \quad
    m_1 = \frac{Y_i(1)-\theta_{10}-\theta_{11}P_i(1)}{\theta_{12}+\theta_{13}P_i(1)}, \quad
    v_1 = \frac{e^{\lambda_0+\lambda_1 P_i(1)}}{\{\theta_{12}+\theta_{13}P_i(1)\}^2}.
\end{equation*}
While for each $i$ such that the observed treatment level is $T=0$, $P_i(1-t)$ is drown from 

\begin{equation*}
    \{P_i^{mis}| S_i^{(1-t)}=l, \eta,\sigma^2, P_i, Y_i\} \sim \mathcal{N}(\eta_l,\sigma^2_l).
\end{equation*}

\vspace{0.25cm}
\noindent {\em Outcome Model.} The $\theta^{(t)}$ parameters are independent for the treatment level $t$, therefor the posterior distributions are, respectively for $t=0,1$:
\begin{align*}
    \theta^{(t)} & \sim \Norm_{q^{(t)}}((V^{(t)})^{-1}M^{(t)},V^{(t)})^{-1}); \\
    V^{(t)} &= (\Tilde{P}^{(t)})^T\Phi^{(t)}\Tilde{P}^{(t)} +(\sigma^2_{\theta})^{-1}I_{q^{(t)}}; \\
     M^{(t)} &= (\Tilde{P}^{(t)})^T\Phi^{(t)}Y(t) +\frac{\mu_{\theta}}{\sigma^2_{\theta}}.
\end{align*}
For the treatment level $t=0$: $q^{(0)}=2$; $\Tilde{P}^{(0)}$ is a matrix $n_0\times n_0$ such that $\Tilde{P}^{(0)}=[1_{n_0},P(0)]$ with $1_{n_0}$ a vector of $1$ and $P(0)$ the vector of observed values of post-treatment variable for the units $n_0$ assigned at the control group, i.e. $t=0$; and $\Phi^{(0)}$ is a diagonal matrix $n_0\times n_0$ with value $\exp(\lambda_0)$ in the diagonal. In similar way, for the treatment level $t=1$: $q^{(1)}=4$; $\Tilde{P}^{(1)}$ is a matrix $n_1\times n_1$ such that $\Tilde{P}^{(1)}=[1_{n_0},P(1),P(0),P(1)\cdot P(0)]$ with $1_{n_1}$ a vector of $1$, $P(1)$ the vector of observed values of post-treatment variable for the units $n_1$ assigned at the treated group, i.e. $t=0$, and $P(0)$ the vector of imputed values of post-treatment variable; and $\Phi^{(1)}$ is a diagonal matrix $n_1\times n_1$ with the values $\exp(\lambda_0+\lambda_1 P(1))$ in the diagonal. 

The parameters in the variance of the $Y$-model, $\lambda_0$ and $\lambda_1$, do not have conjugate priors, therefore a independent Metropolis proposal step is necessary. At each iteration $r\in \{1,\dots,R\}$, $\lambda_0^*$ and $\lambda_1^*$ are drown from the proposal distribution $\mathcal{N}(\mu_{\lambda_0},\sigma^2_{\lambda_0})$ and $\mathcal{N}(\mu_{\lambda_1},\sigma^2_{\lambda_1})$ respectively. Then, at iteration $r$ the value of the parameter are updated as following: $\lambda_0^{(r)}=\lambda_0^*$ with probability
\begin{equation*}
    \frac{\prod_{i\in n} \Norm(Y_i|\mu_{Y}^{(t)},\exp(\lambda_0^{*}+\iv_{(T_i=1)}\lambda_1^{(r-1)} P_i(1))}{\prod_{i\in n} \Norm(Y_i|\mu_{Y}^{(t)},\exp(\lambda_0^{(r-1)}+\iv_{(T_i=1)}\lambda_1^{(r-1)} P_i(1))},
\end{equation*}
otherwise $\lambda_0^{(r)}=\lambda_0^{(r-1)}$; and  $\lambda_1^{(r)}=\lambda_1^*$ with probability
\begin{equation*}
     \frac{\prod_{i\in n_1} \Norm(Y_i|\mu_{Y}^{(t)},\exp(\lambda_0^{(r-1)}+\lambda_1^{*} P_i(1))}{\prod_{i\in n_1} \Norm(Y_i|\mu_{Y}^{(t)},\exp(\lambda_0^{(r-1)}+\lambda_1^{(r-1)} P_i(1))},
\end{equation*}
otherwise $\lambda_1^{(r)}=\lambda_1^{(r-1)}$; where $\mu_Y^{(0)}=\theta_{00} + \theta_{01}P_i(0)$ and $\mu_Y^{(1)}=\theta_{10} + \theta_{11}P_i(1)+ \theta_{12}P_i(0) + \theta_{13}P_i(0)P_i(1)$. 

{\vspace{0.8cm}
%\spacingset{1}
\begin{algorithm}
\caption{Confounders-Aware Shared-atoms Bayesian Mixture Model }
\label{alg:common}
\vspace{0.15cm}
%\hrule
{\bf Inputs:} 

\quad - the observed data $(y,p,t,x)$.

{\bf Outputs:} 

\quad - posterior distributions of parameters: $\eta$, $\sigma$, $\beta$, $\theta$, and $\lambda$;

\quad - imputed values for $P^{mis}$;

\quad -  posterior distribution over the space of partitions of the units.
\vspace{0.05cm}

%\hrule
{\bf Procedure:}\\
Initialization of all parameters and latent variables.\\
\textbf{For} $r \in \{1,\dots,R\}$ \textbf{:}

\quad $\xrightarrow{}$ \textit{Estimation of Shared Atoms Mixture Model:}

\qquad\quad Compute $\omega^{(t)}(x_i)$ for $i=1,\dots,n$ and $t= 0,1$;

\qquad\quad Draw $S_{i}^{(t)}$ for $i=1,\dots,n$ and $t= 0,1$;

\qquad\quad Draw $\eta$ and  $\sigma$;

\qquad\quad Compute $\alpha^{(t)}(x_i)$ for $i=1,\dots,n$ and $t= 0,1$;

\qquad\quad Draw $\beta^{(t)}$ for$t= 0,1$.

\quad $\xrightarrow{}$ \textit{Imputation of Missing Post-Treatment Variables:}

\qquad\quad Draw $P_i^{mis}$ for $i=1,\dots,n$ and $t= 0,1$.

\quad $\xrightarrow{}$ \textit{Estimation of Outcome Model:}

\qquad\quad Draw $\theta^{(t)}$ for $t= 0,1$;

\qquad\quad Draw $\lambda_0$ and $\lambda_1$.\\
\textbf{End} 
%\hrule
\end{algorithm}}

%% file: Appendix/appendix_simulations.tex
%Additionally to the three scenarios reported in the Section~\ref{sec:simulation}, we examine two scenarios where the groups have the atoms for the post-treatment variables are closer, and the variances are bigger and different among the groups. The parameters for the Y-model are the same in the Section~\ref{sec: simulation}.

%\vspace{0.25cm}

\begin{figure}[!htp]
    \centering
    \includegraphics[width=0.3\linewidth]{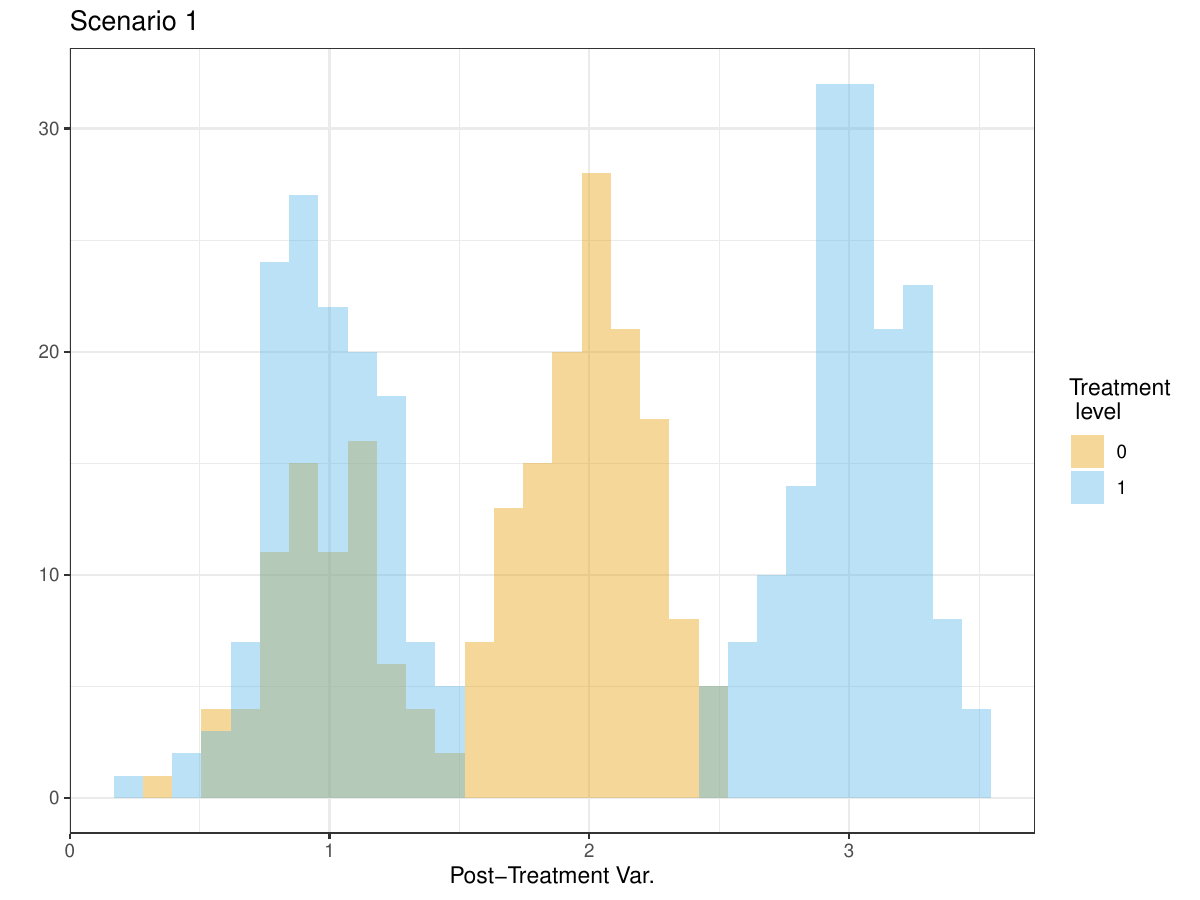}\;  \includegraphics[width=0.3\linewidth]{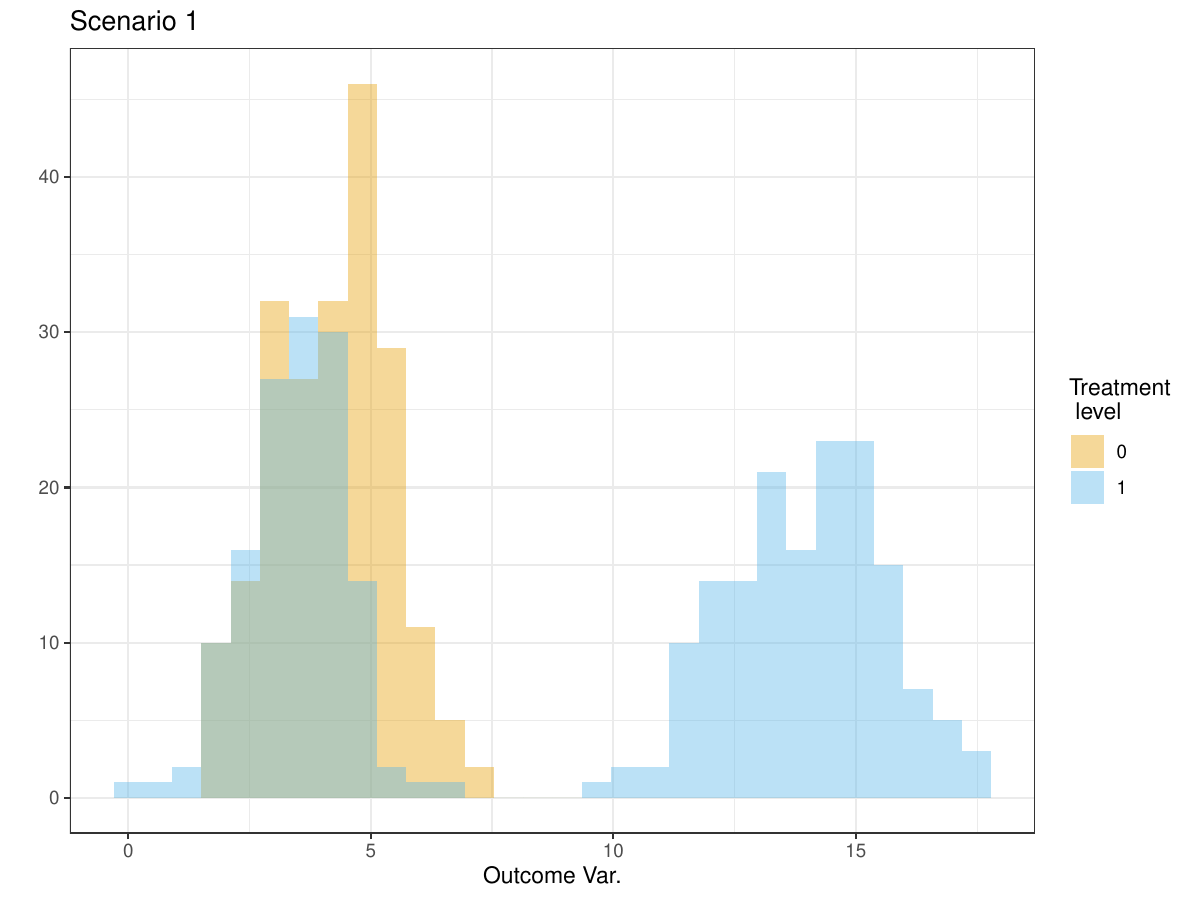} \\
    \includegraphics[width=0.3\linewidth]{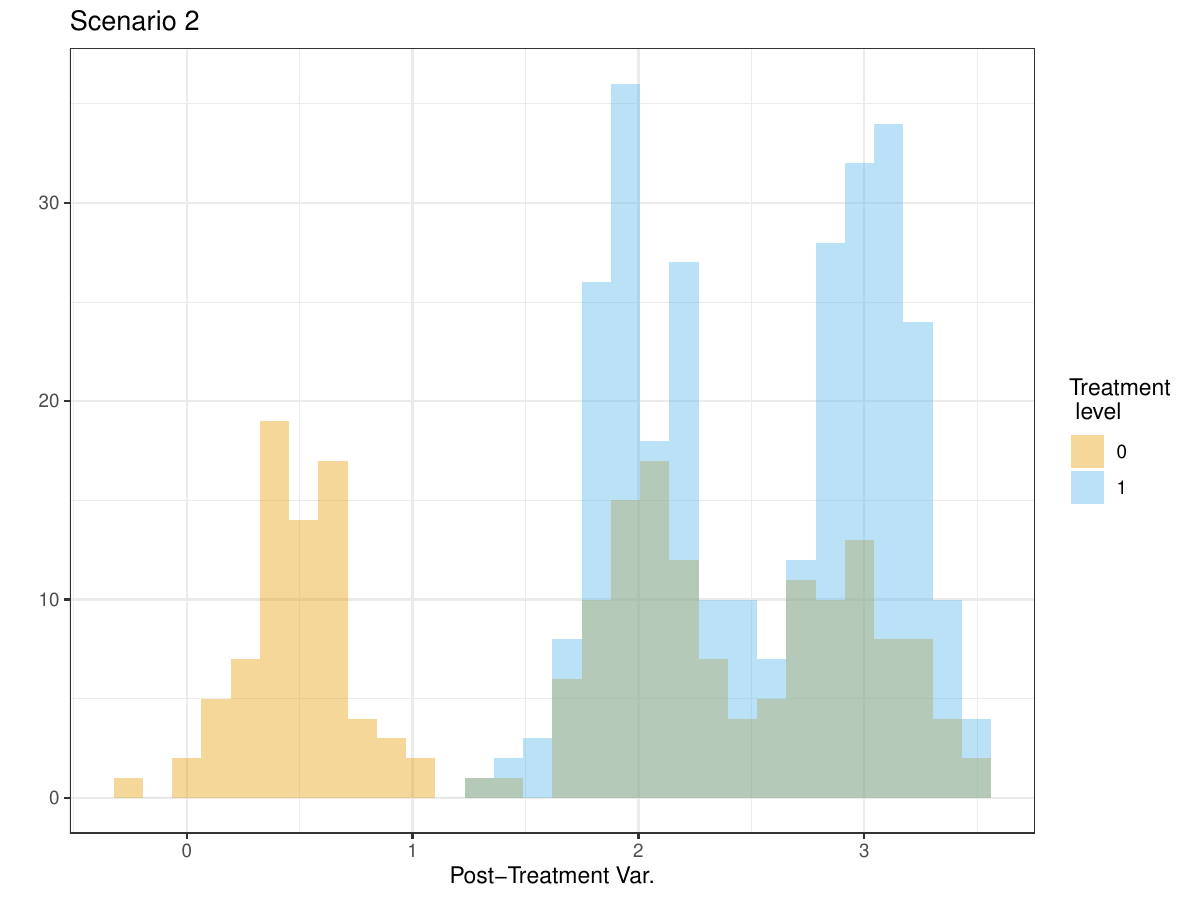}\;  \includegraphics[width=0.3\linewidth]{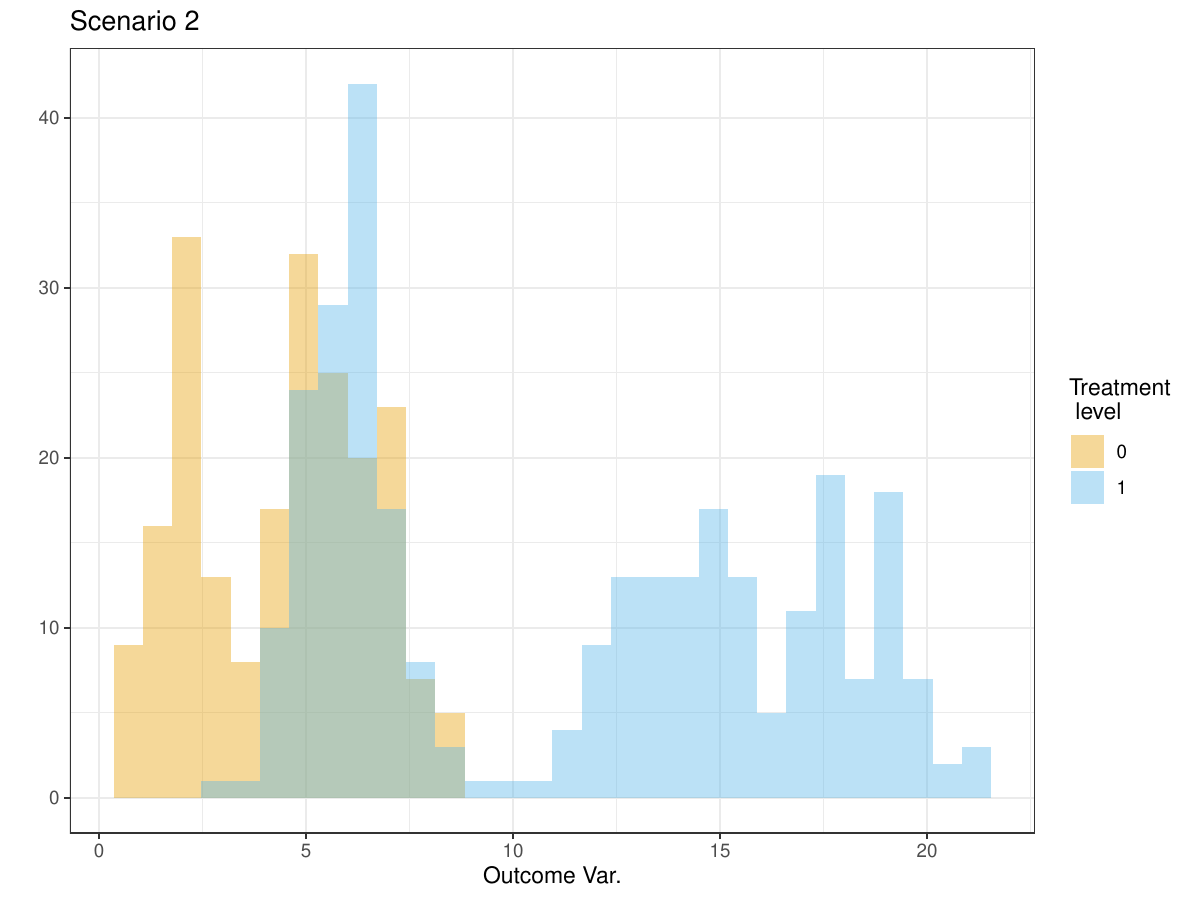} \\
    \includegraphics[width=0.3\linewidth]{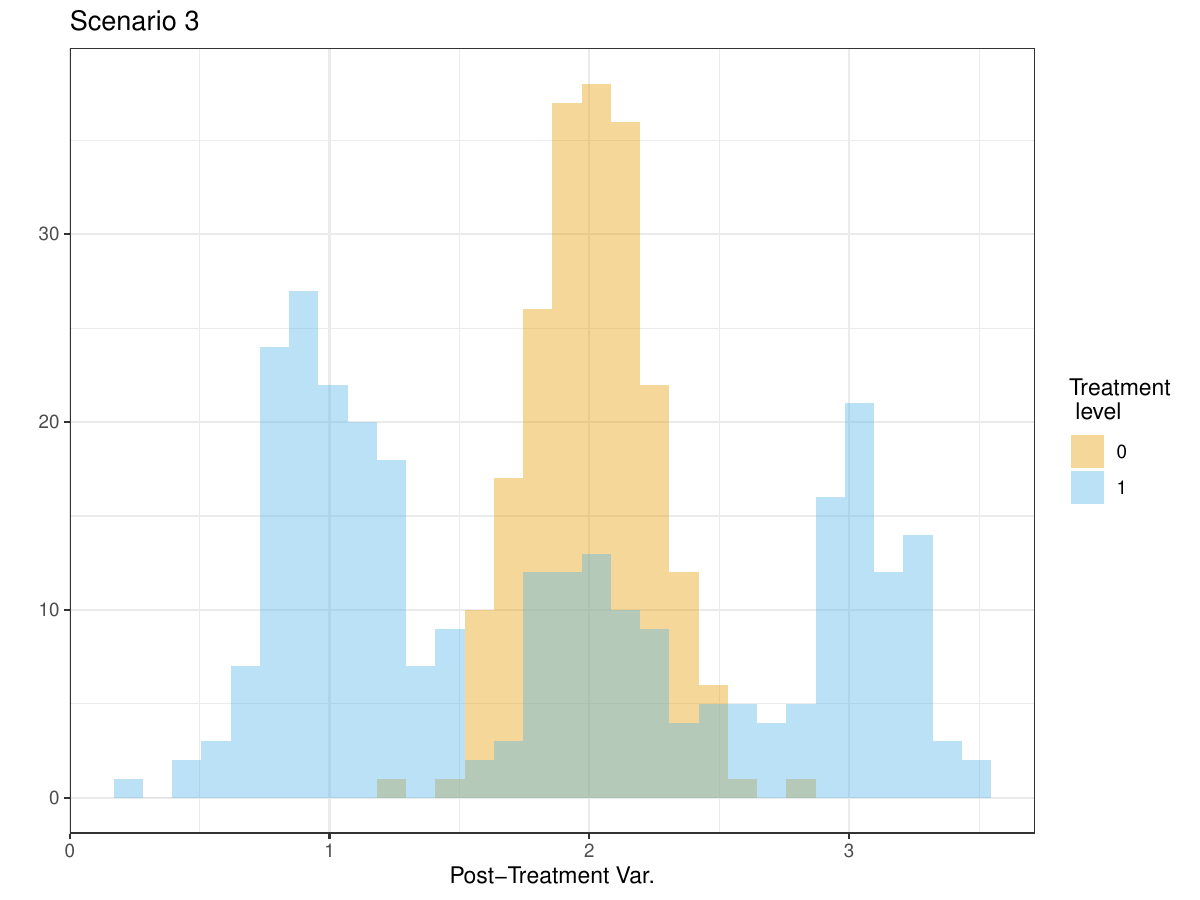}\;  \includegraphics[width=0.3\linewidth]{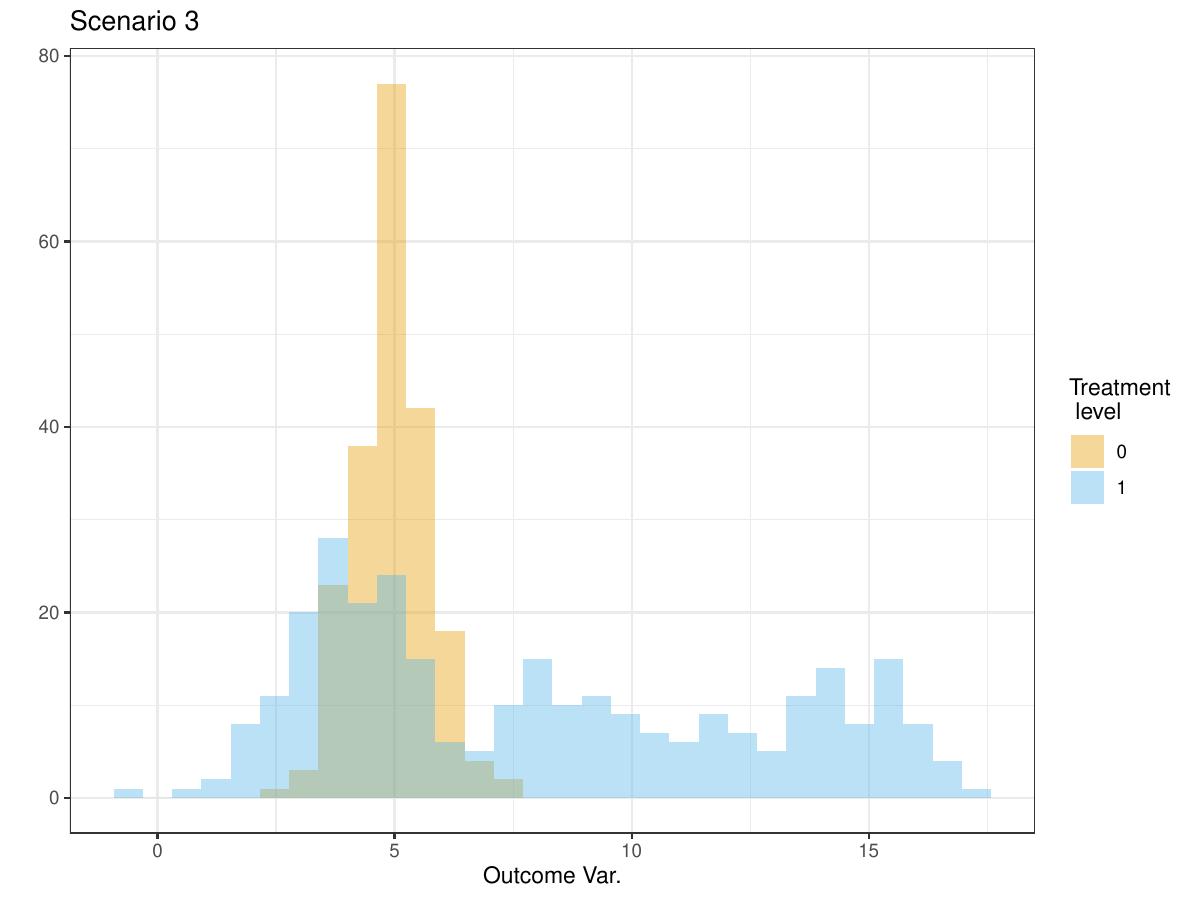} \\
    \includegraphics[width=0.3\linewidth]{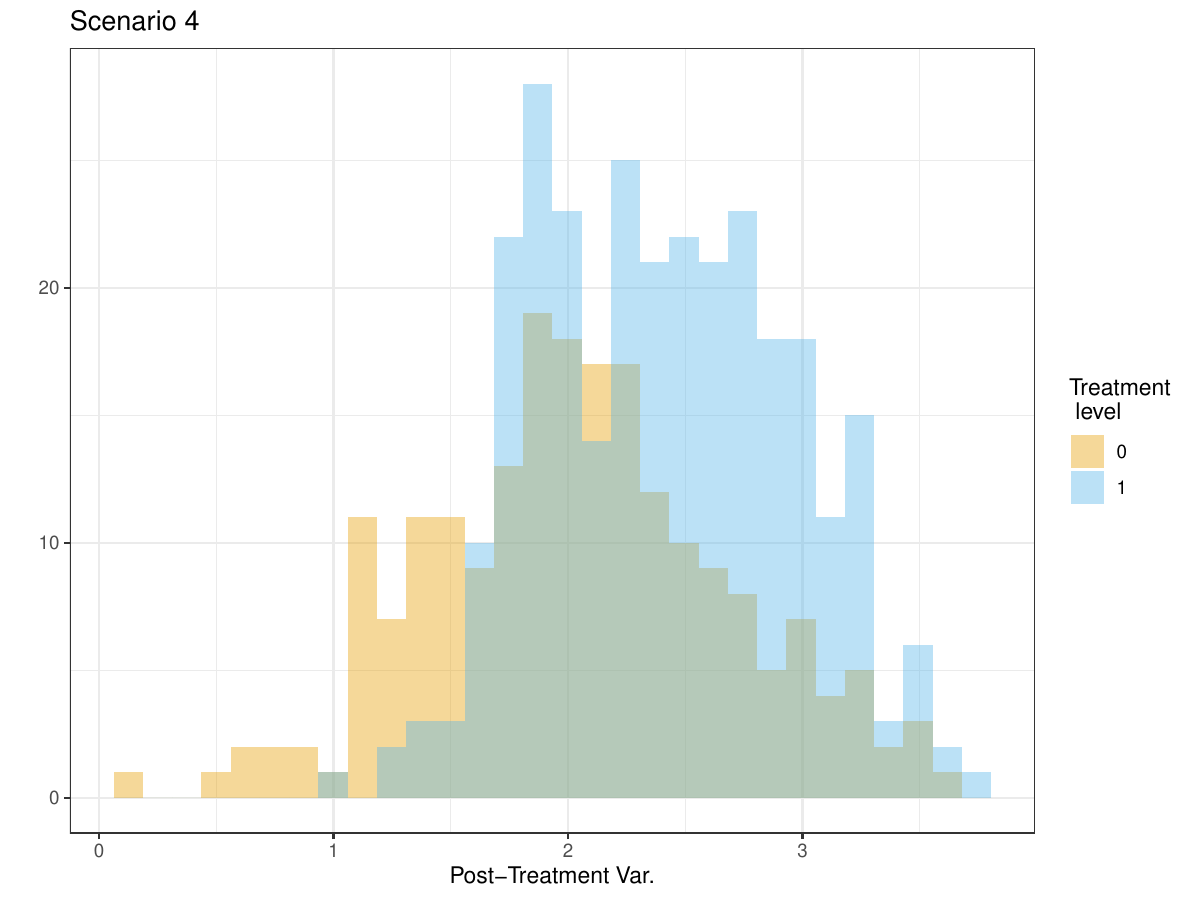}\;  \includegraphics[width=0.3\linewidth]{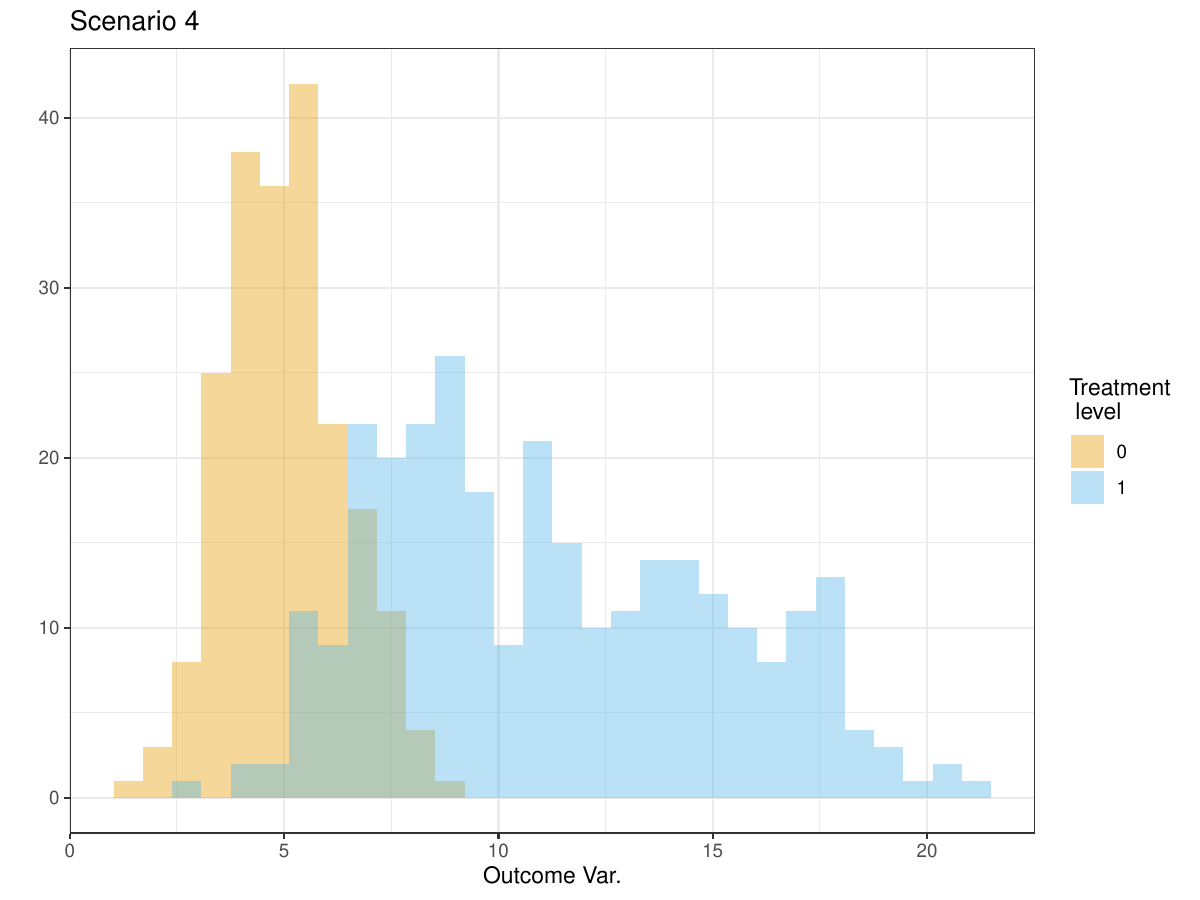} \\
    \includegraphics[width=0.3\linewidth]{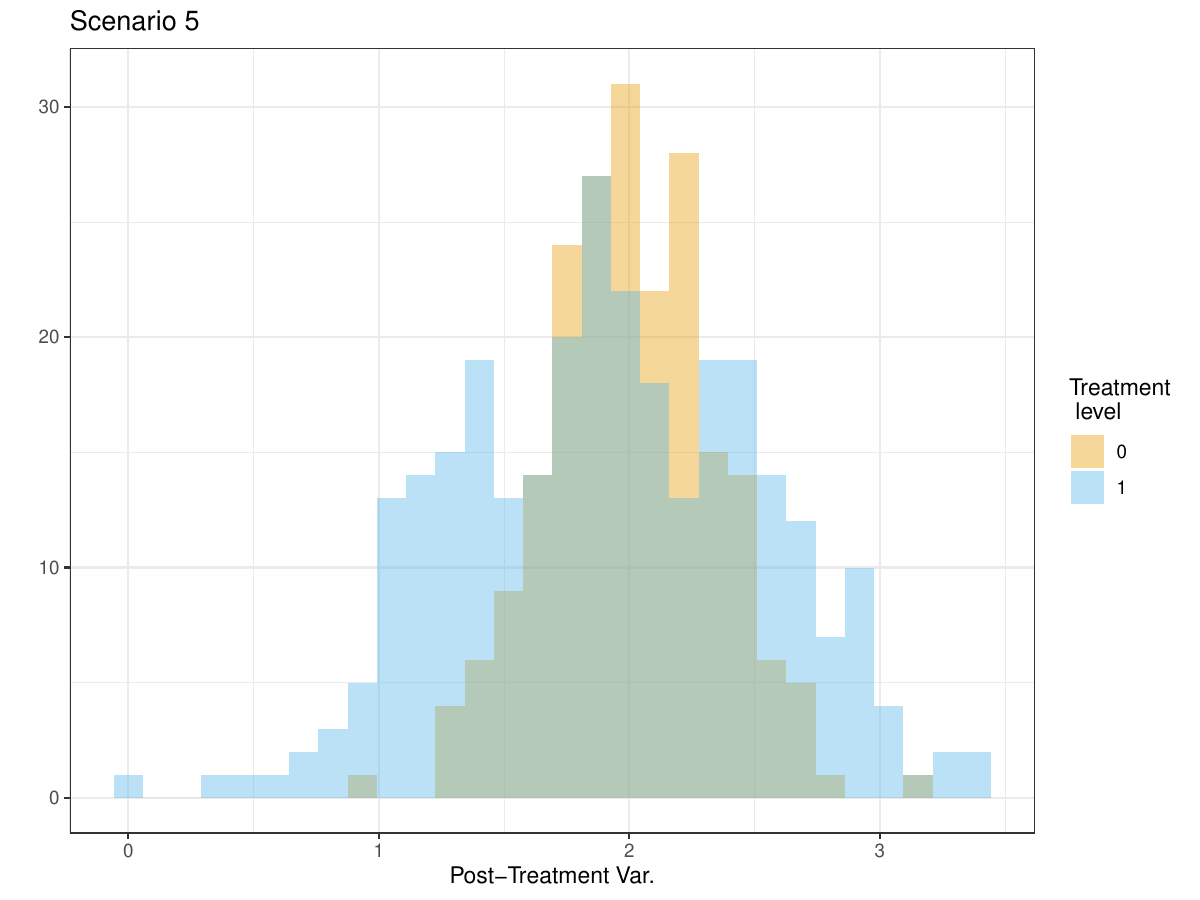}\;  \includegraphics[width=0.3\linewidth]{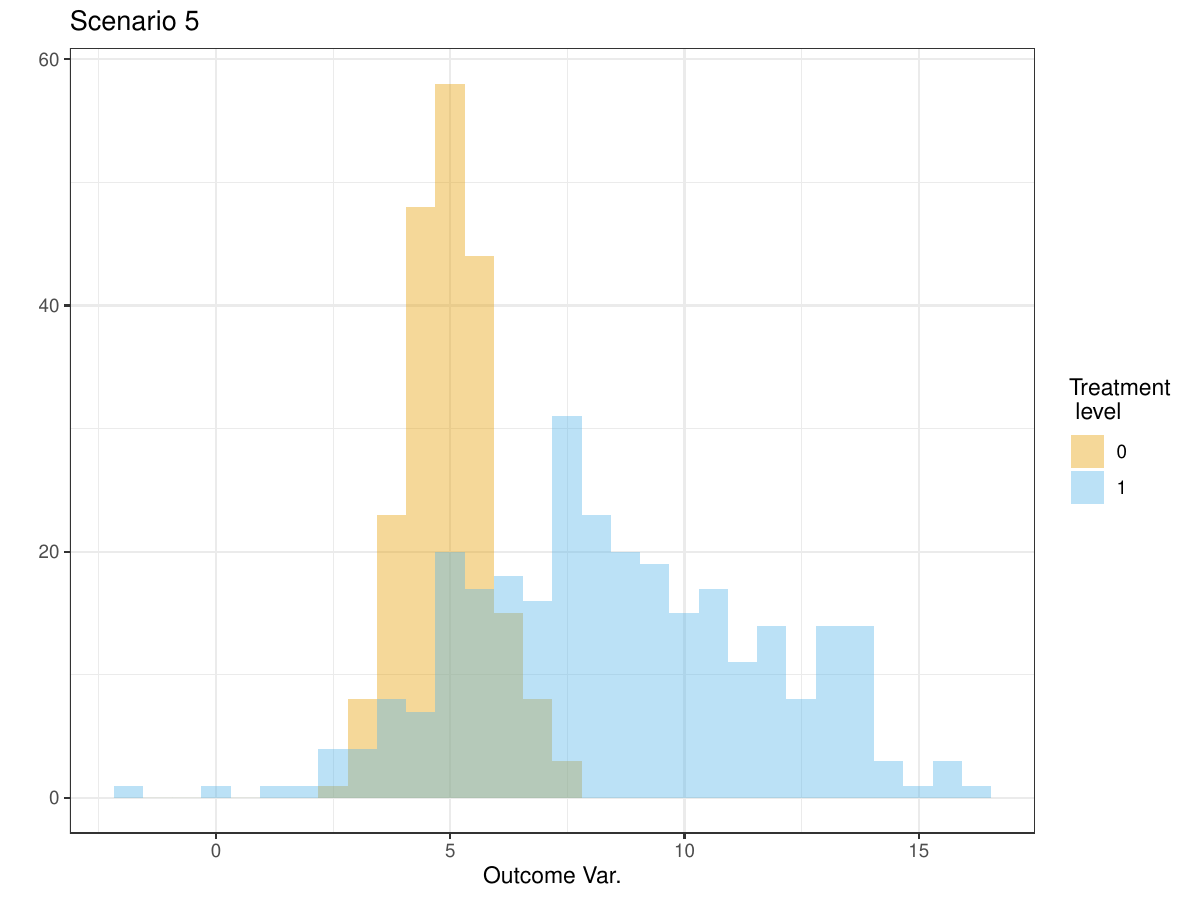} 
\caption{Distributions of (right) the post-treatment variables and (left) the outcomes, for the five simulated scenarios. In light blue the distribution of the variables given the treatment and in yellow given the control.}
\label{Figure:simulations_appendix}
\end{figure}

Figure~\ref{Figure:simulations_appendix} reports the distribution of the simulated post-treatment variables and outcomes for the five simulated scenarios. The complexity of the distributions increases over the simulated scenarios, in particular in scenarios 4 and 5, it is not unidentifiable the groups in the distributions of the potential post-treatment variables and the potential outcomes. 

\begin{figure}[!htp]
\begin{center}
\includegraphics[width=2.1in]{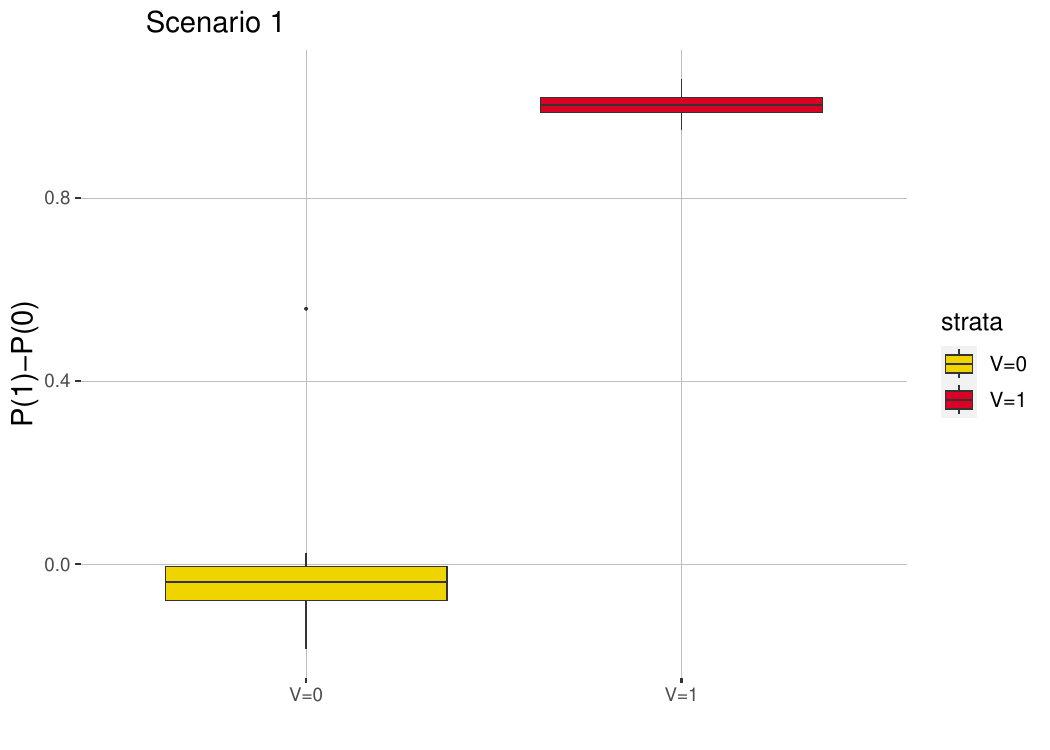}\; \includegraphics[width=2.1in]{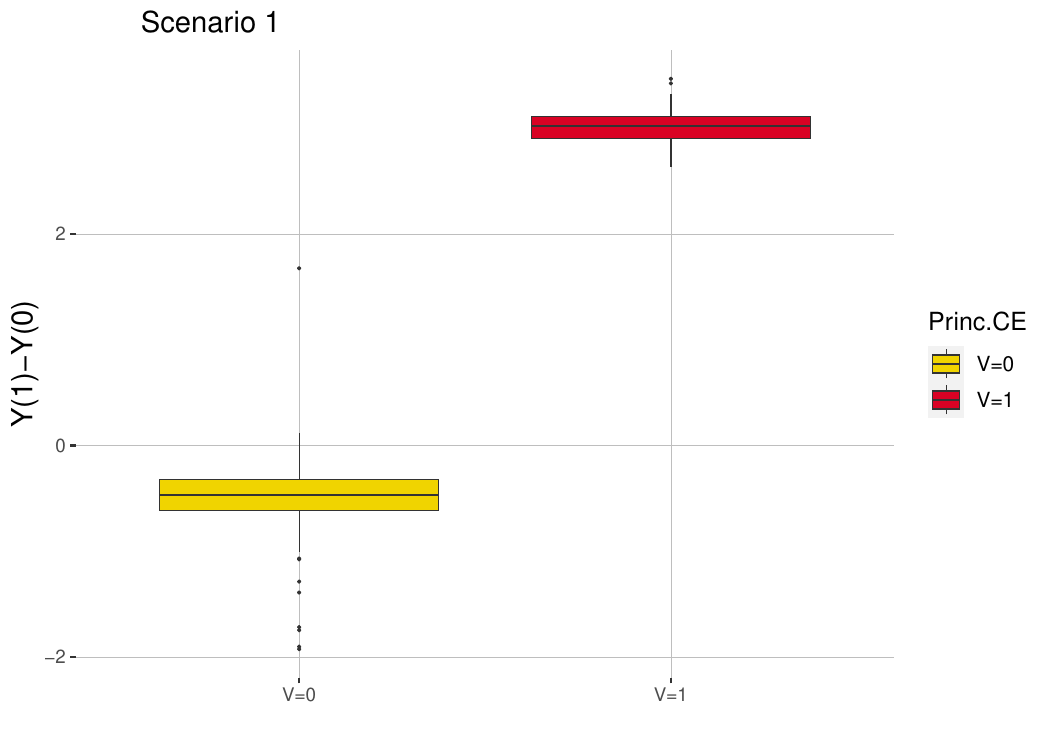}\\
%\vspace{0.3em}
\includegraphics[width=2.1in]{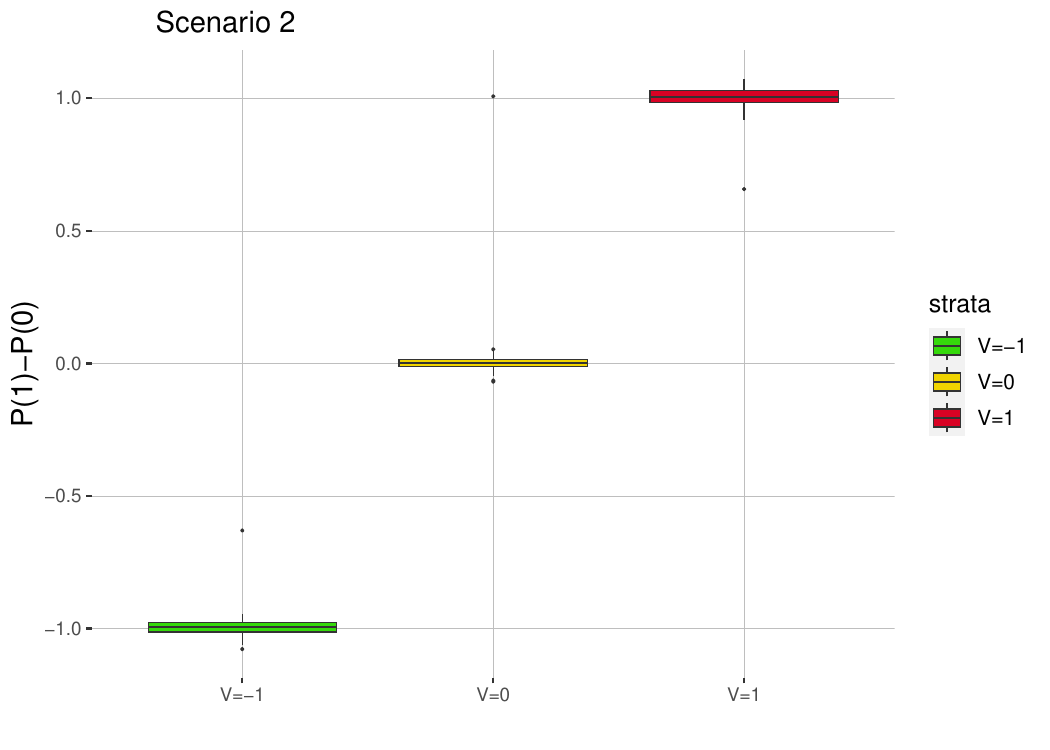}\; \includegraphics[width=2.1in]{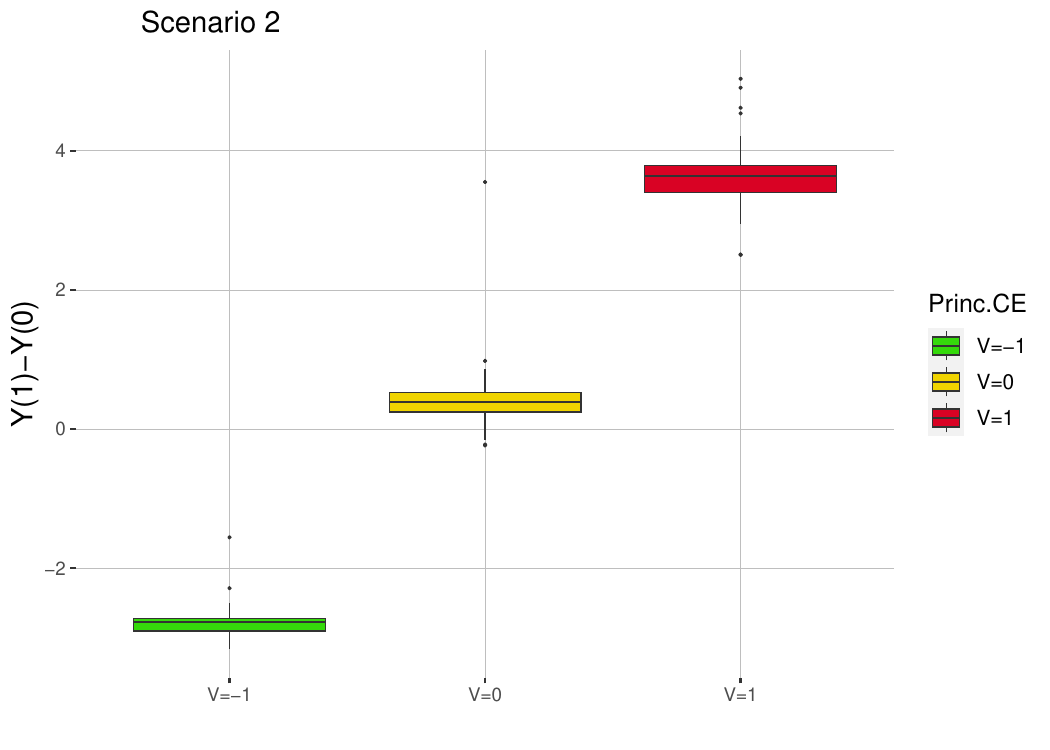}\\
%\vspace{0.3em}
\includegraphics[width=2.1in]{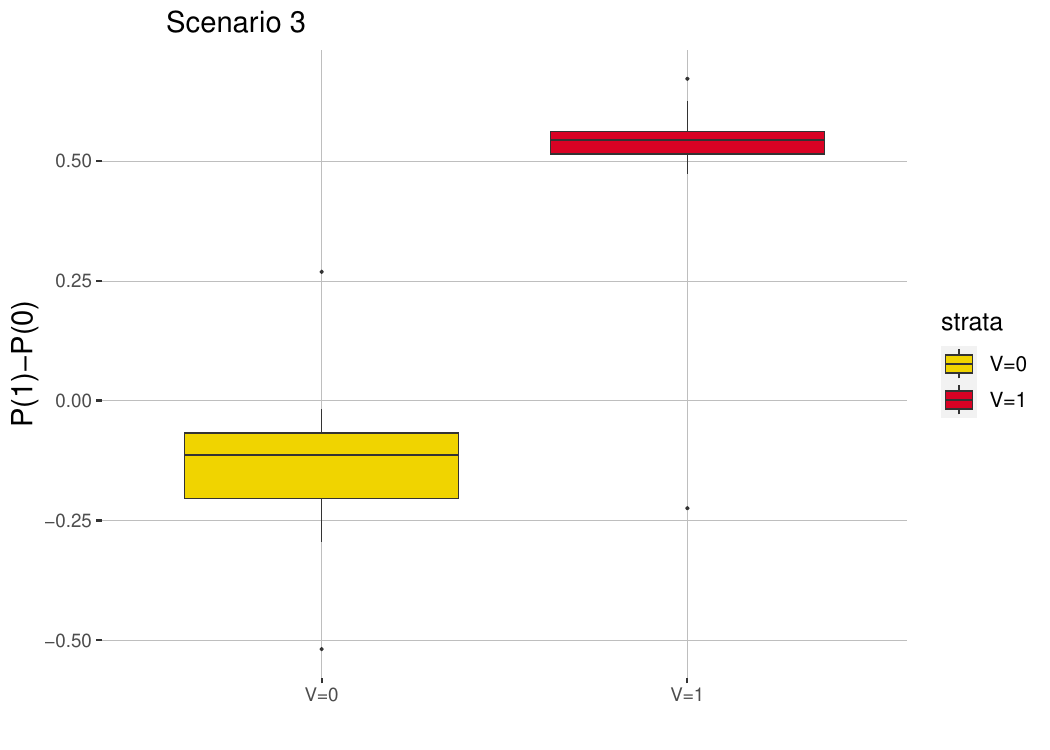}\; \includegraphics[width=2.1in]{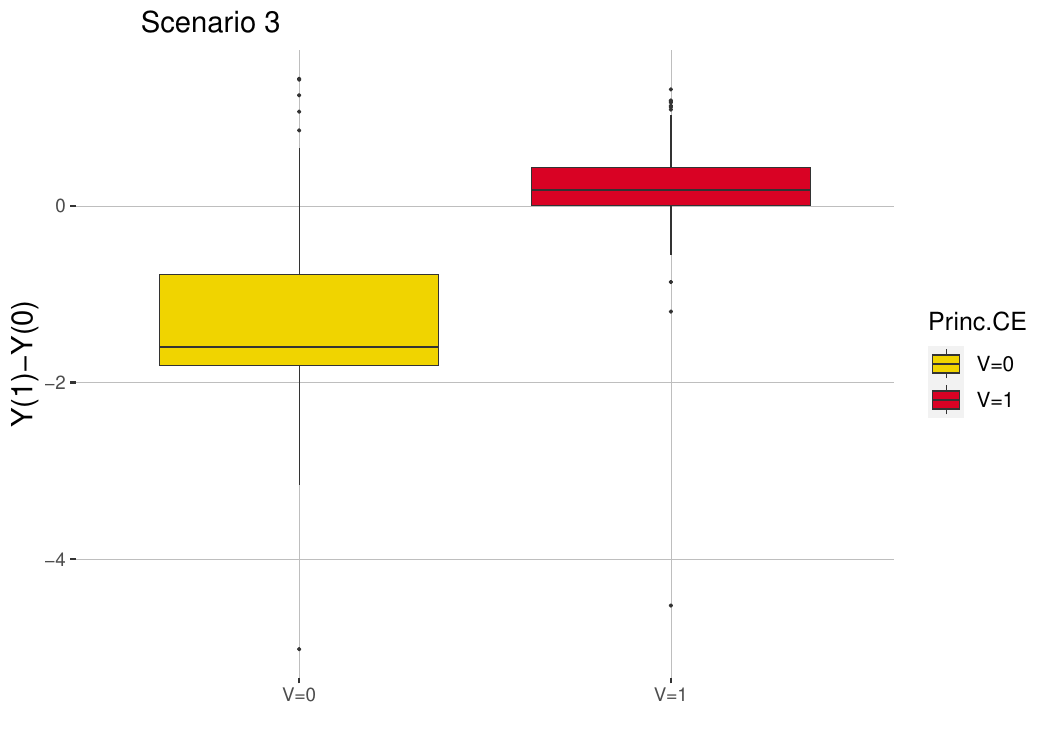}\\
%\vspace{0.3em}
\includegraphics[width=2.1in]{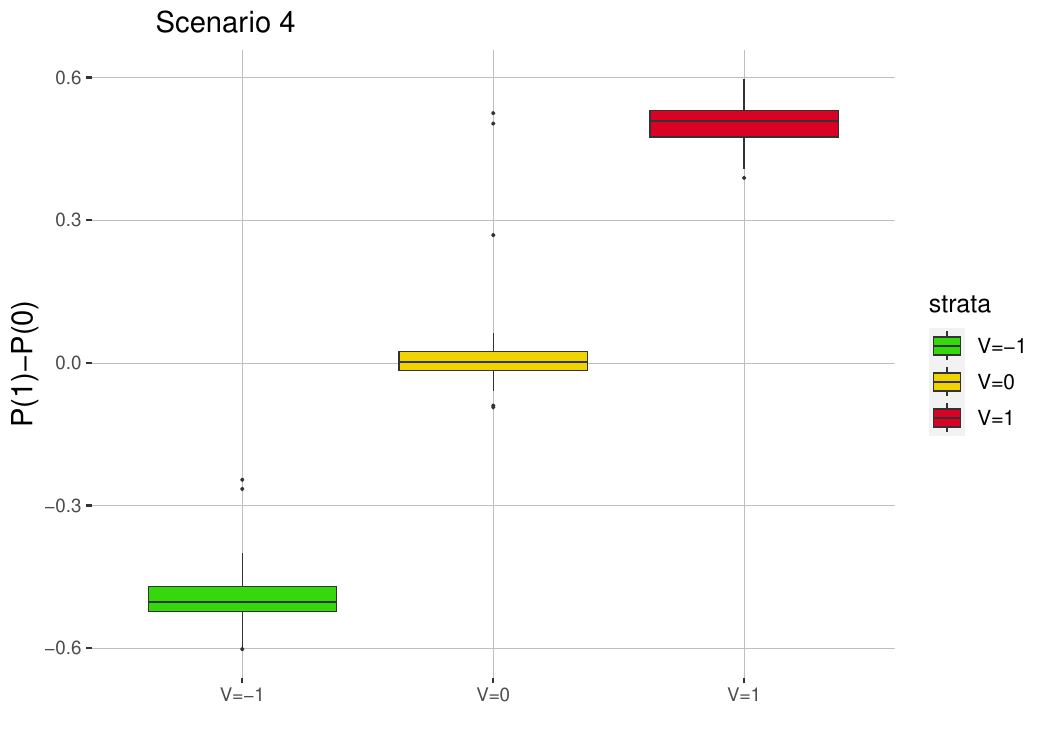}\; \includegraphics[width=2.1in]{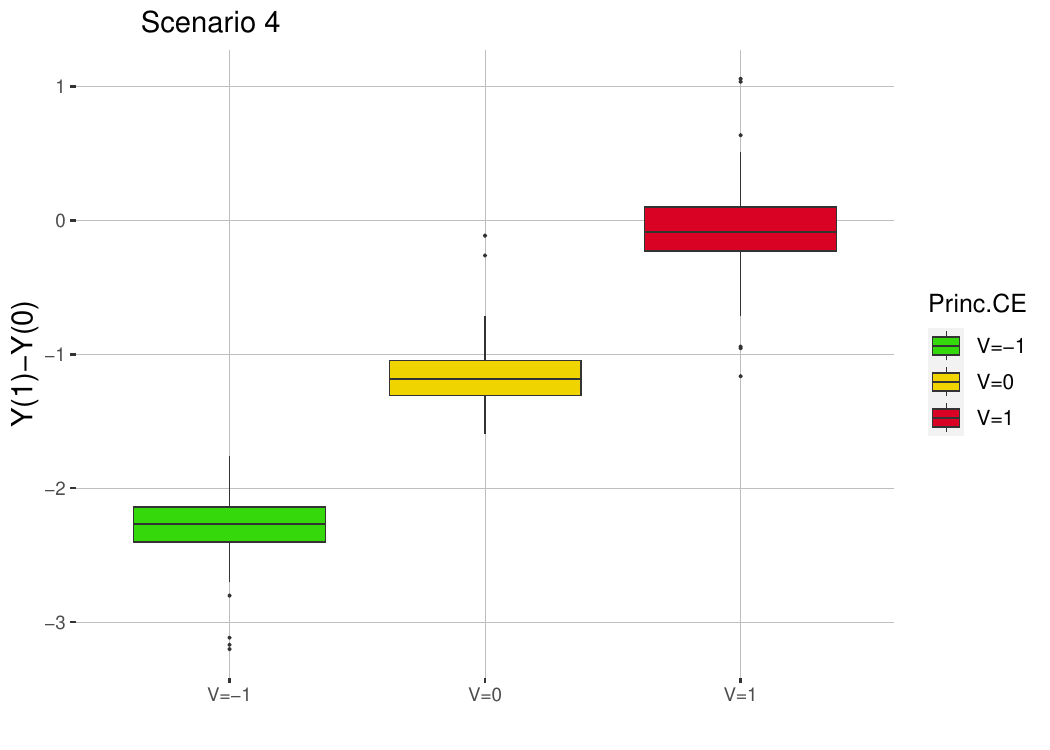}\\
%\vspace{0.3em}
\includegraphics[width=2.1in]{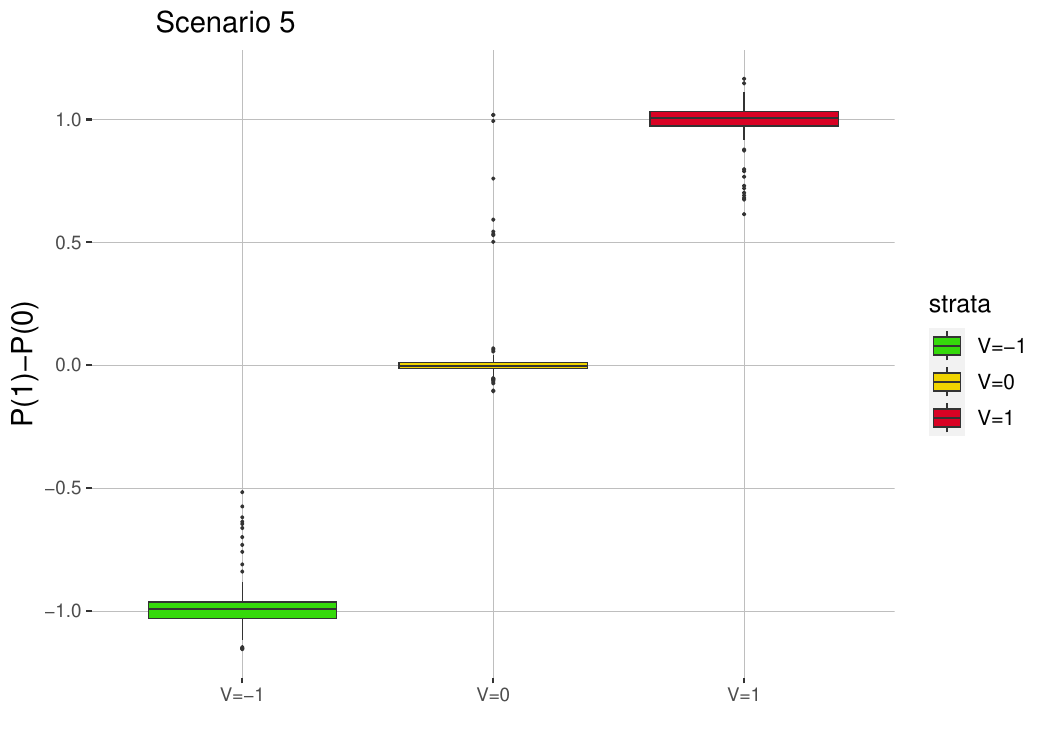}\; \includegraphics[width=2.1in]{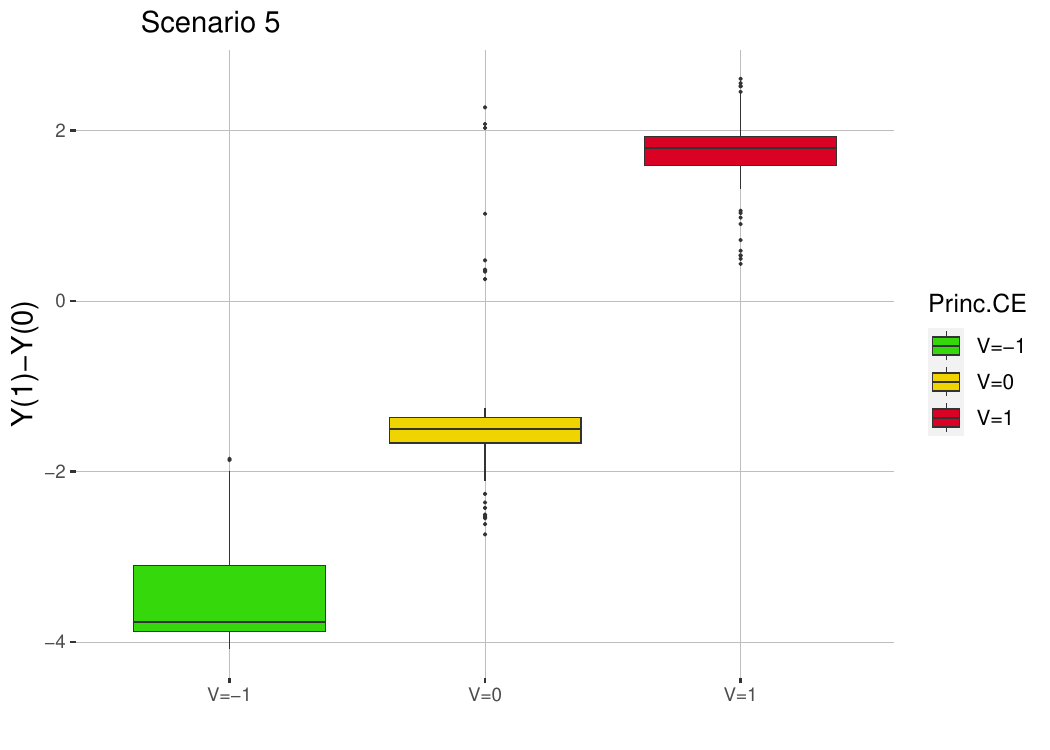}
\end{center}
\caption{Representations of the five simulated scenarios. (Left) The expected value of the difference of the post-treatment variables under treatment and control given the strata allocation. (Right) The expected associative/dissociative causal effects. In green and indicated with $V=-1$, the associative negative stratum, i.e. corresponding to $S_i^{(1)} \prec S_i^{(0)}$, in yellow and indicated with $V=0$, the dissociative stratum, i.e., $S_i^{(1)} = S_i (0)$, and in red and indicated with $V=+1$ the associative positive stratum, i.e., $S_i^{(1)} \succ S_i^{(0)}$. \label{fig:sim_strata}}
\end{figure}

Figure~\ref{fig:sim_strata} reports the results for the five simulated scenarios obtained with our proposed model. On the left, the boxplots show the distribution over the simulated samples of the expected values of the difference of the post-treatment variables under treatment and control in each stratum. The graphics confirm the ability of our proposed model to (i) identify correctly the number of strata:two strata in the simulated scenario 1 and 3, and three in the others, and (ii) capture the definition of the associative/dissociative strata without an \textit{a priori} criteria: the dissociative stratum is always around zero for $\E\{P(1)-P(0)\}$, while the dissociative stratum does not include the zero.
In the boxplots on the right, there is the distribution of the principal causal effect: $\tau_-$, $\tau_0$, and $\tau_+$. The different strata identify different treatment effects on the outcome, allowing us to characterize the heterogeneity in the causal effects. Few outliers are observed, however, they are found in particular in Scenario 5 which describes a more complex relation among variables and strata.

%% file: Appendix/appendix_application.tex
\begin{figure}[!htb]
\begin{center}
\includegraphics[width=2.4in]{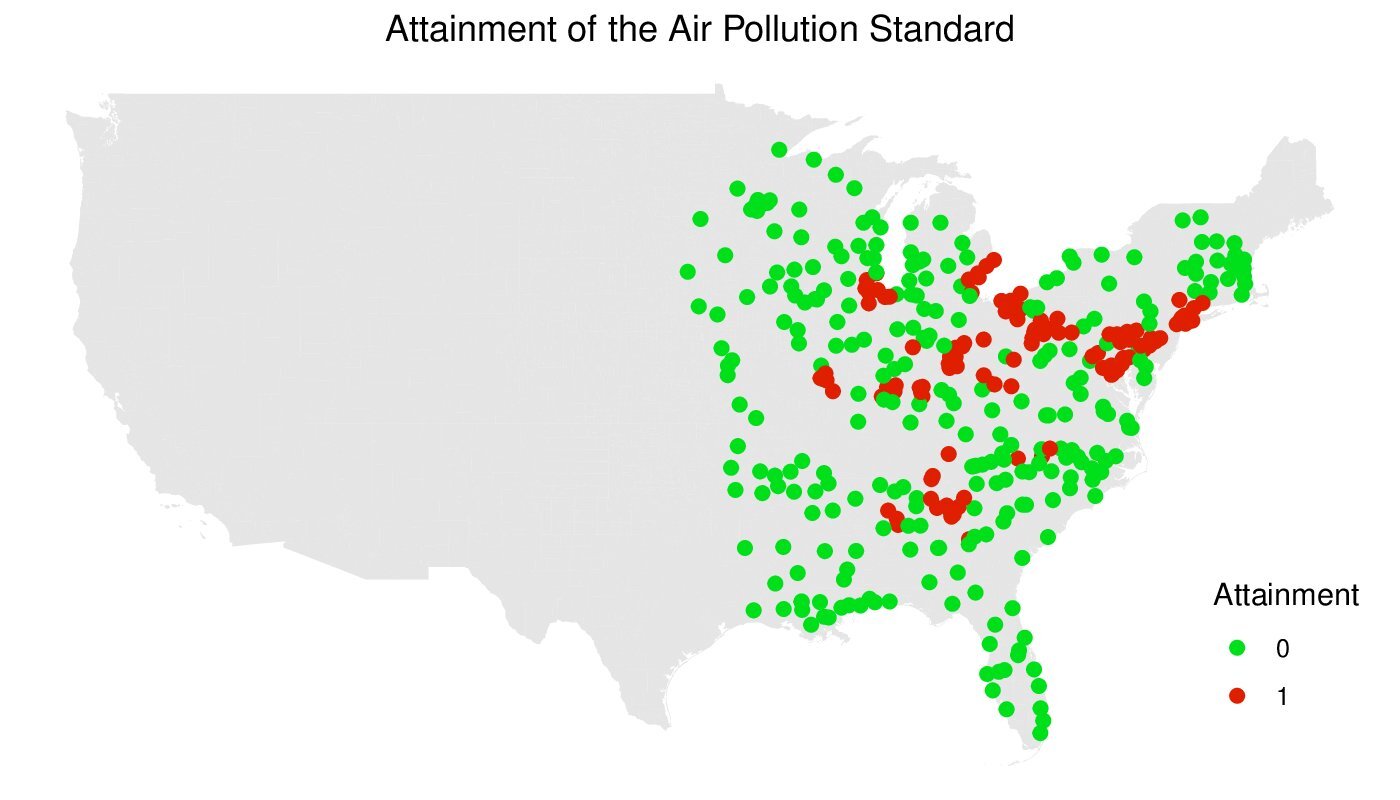}\\
\includegraphics[width=2.4in]{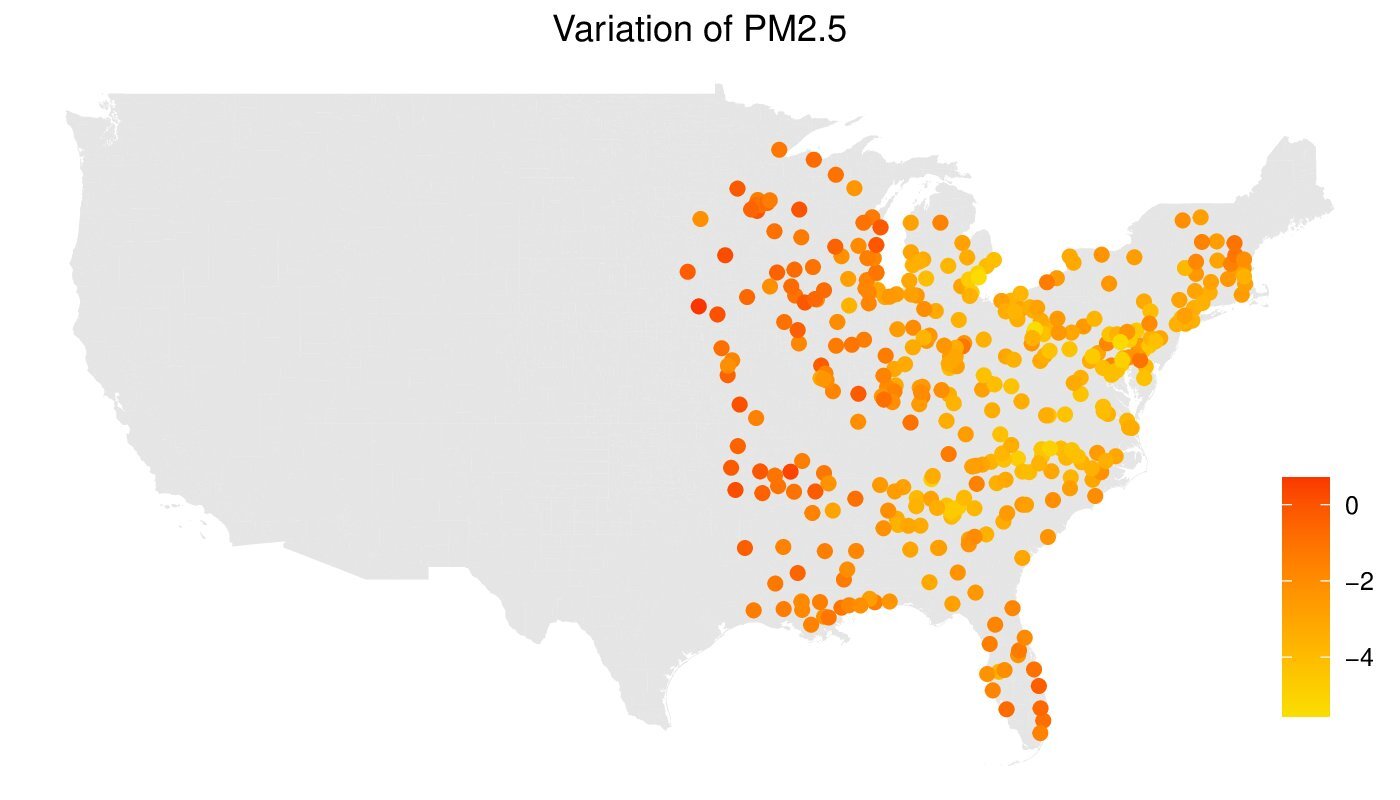}\, \includegraphics[width=2.4in]{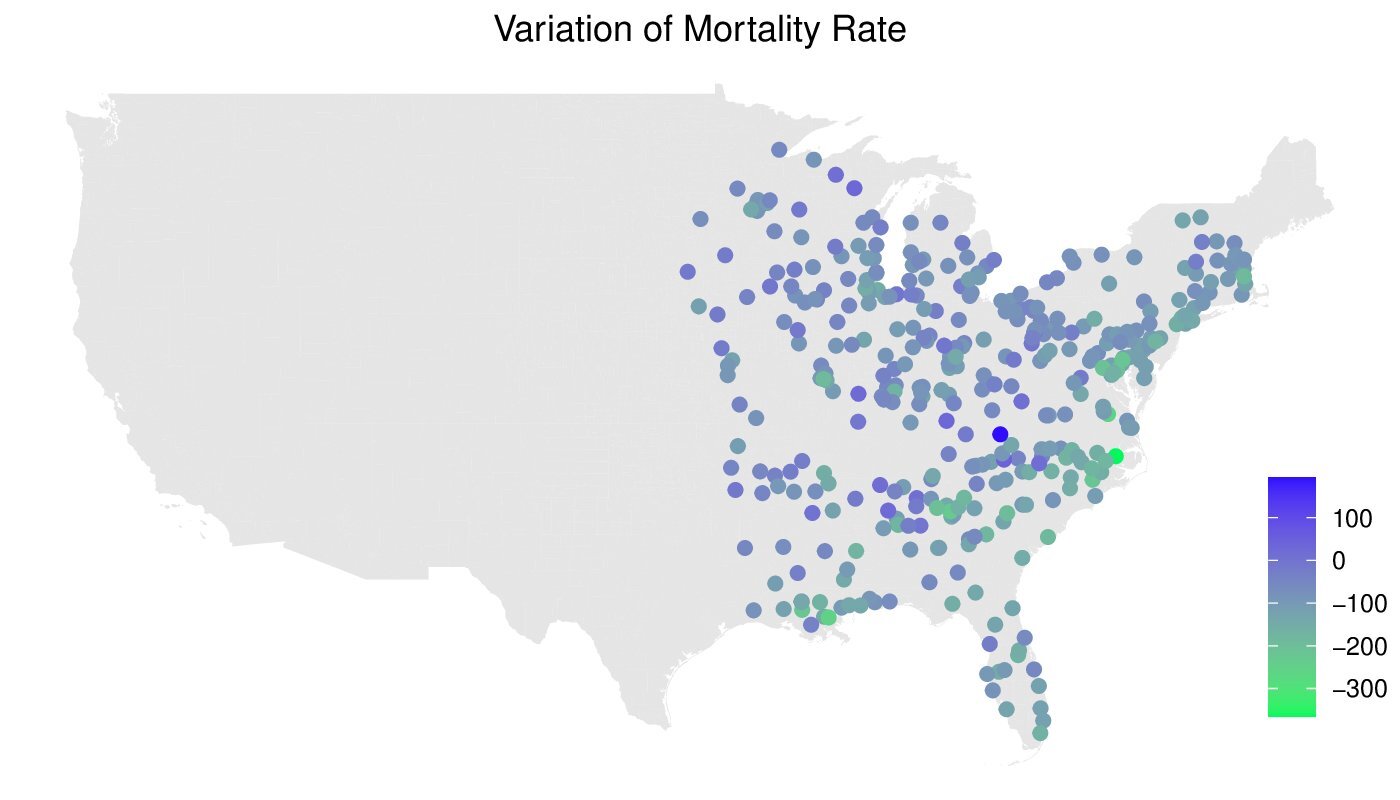}
\end{center}
\caption{Considered counties in the Easter U.S. (top) Attainment of the air pollution standard ($15\mu m$ for \PMns) and consequent application of air pollution regulations: $0$ if the county was in attainment, $1$ if the county was in non-attainment. (bottom left) Difference between the average long-term exposure \PM between baseline and the follow-up period (measured in $\mu m$). (bottom right) Difference of the age-adjusted mortality rate between baseline and the follow-up period, value per 100,000.  \label{fig:map_east_us}}
\end{figure}

Figure~\ref{fig:map_east_us} maps the final analysis dataset described in Section 5.1. On the top map in Figure~\ref{fig:map_east_us}, the red points visualize the treated counties, which appear to be closer to the main cities, such as Chicago, New York City, Washington DC or Cleveland. The continuous post-treatment variable is the difference in level \PM between the follow-up period (2010-2016) and the baseline period (2000-2005), reported on the map to the left of the bottom of Figure~\ref{fig:map_east_us}. The outcome variable is defined as the difference between the age-adjusted mortality rate between the follow-up period and the baseline period, and is visualized in the map at the bottom right of Figure~\ref{fig:map_east_us}. As reported in the maps, both the post-treatment variable and the outcome have almost all negative values, highlighting a general trend to decrease the level of \PM and a reduction in mortality rates in the last decade.

\begin{figure}[!htb]
\centering
\includegraphics[width=2.5in]{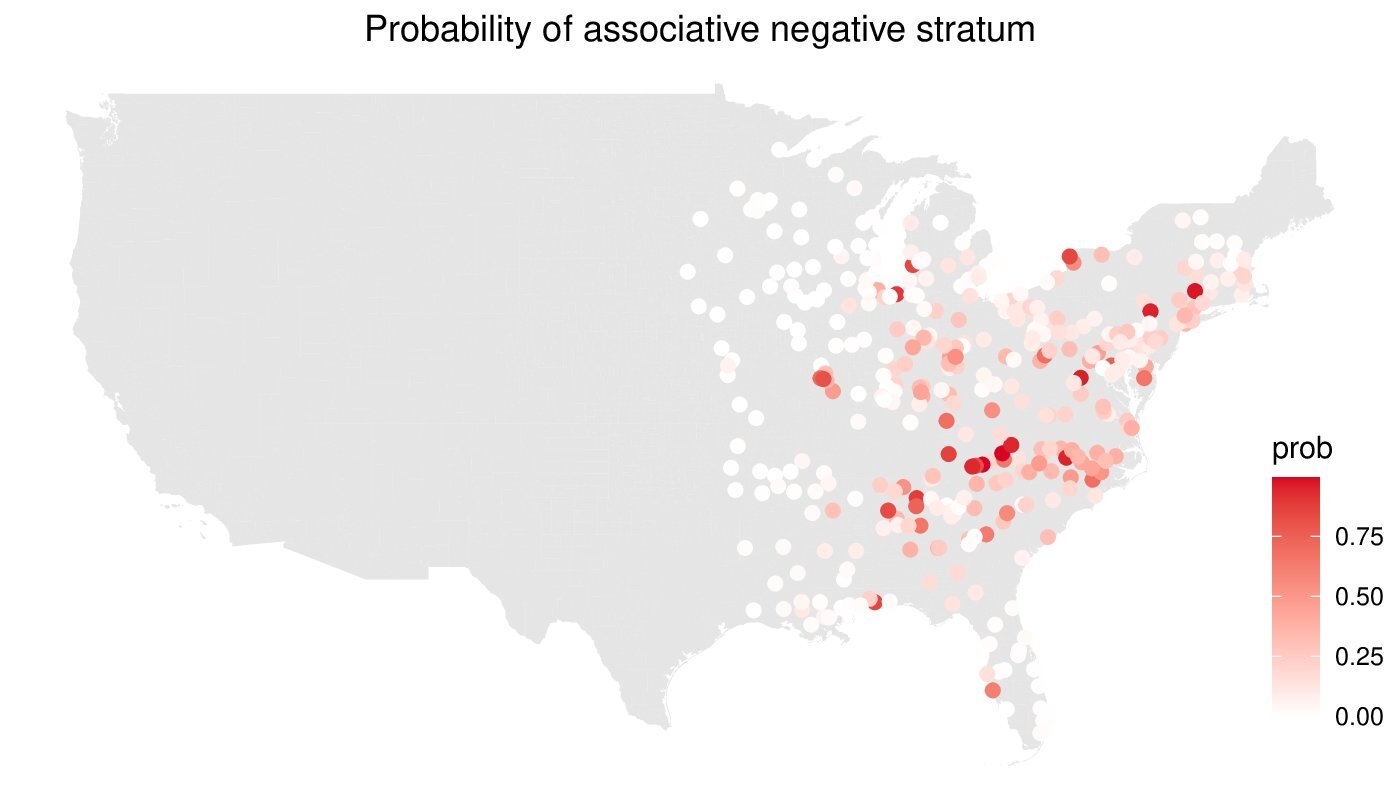}\;
\includegraphics[width=2.5in]{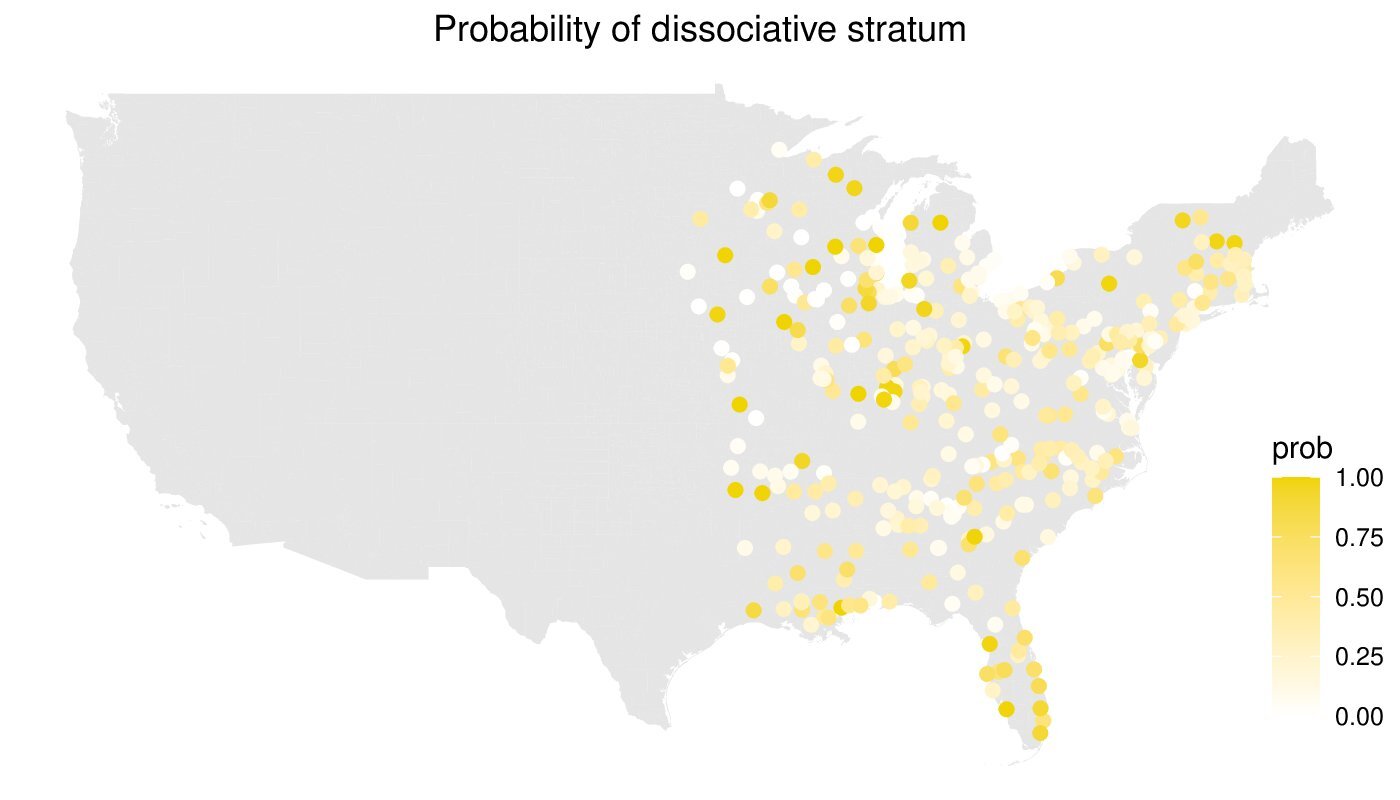}\\ \includegraphics[width=2.5in]{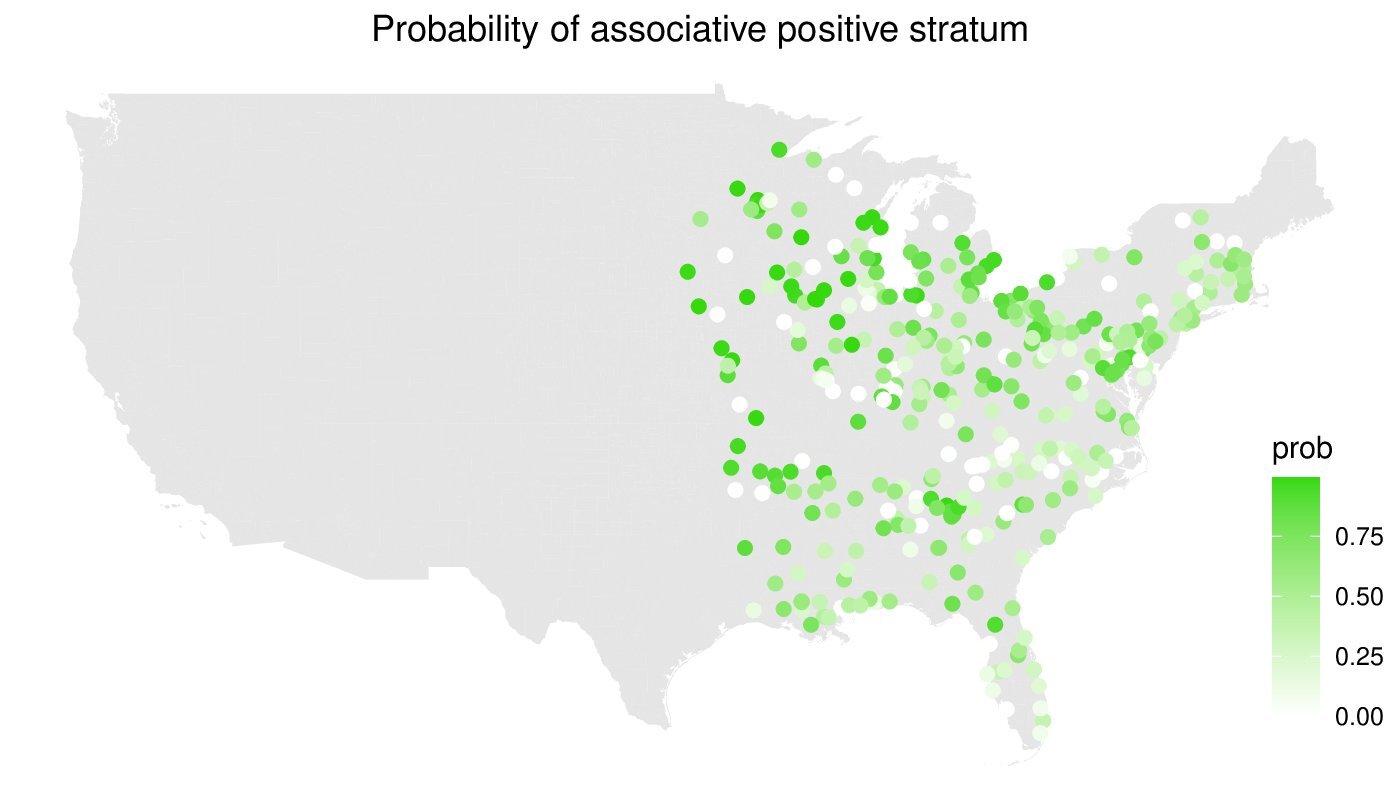}\;
\includegraphics[width=2.5in]{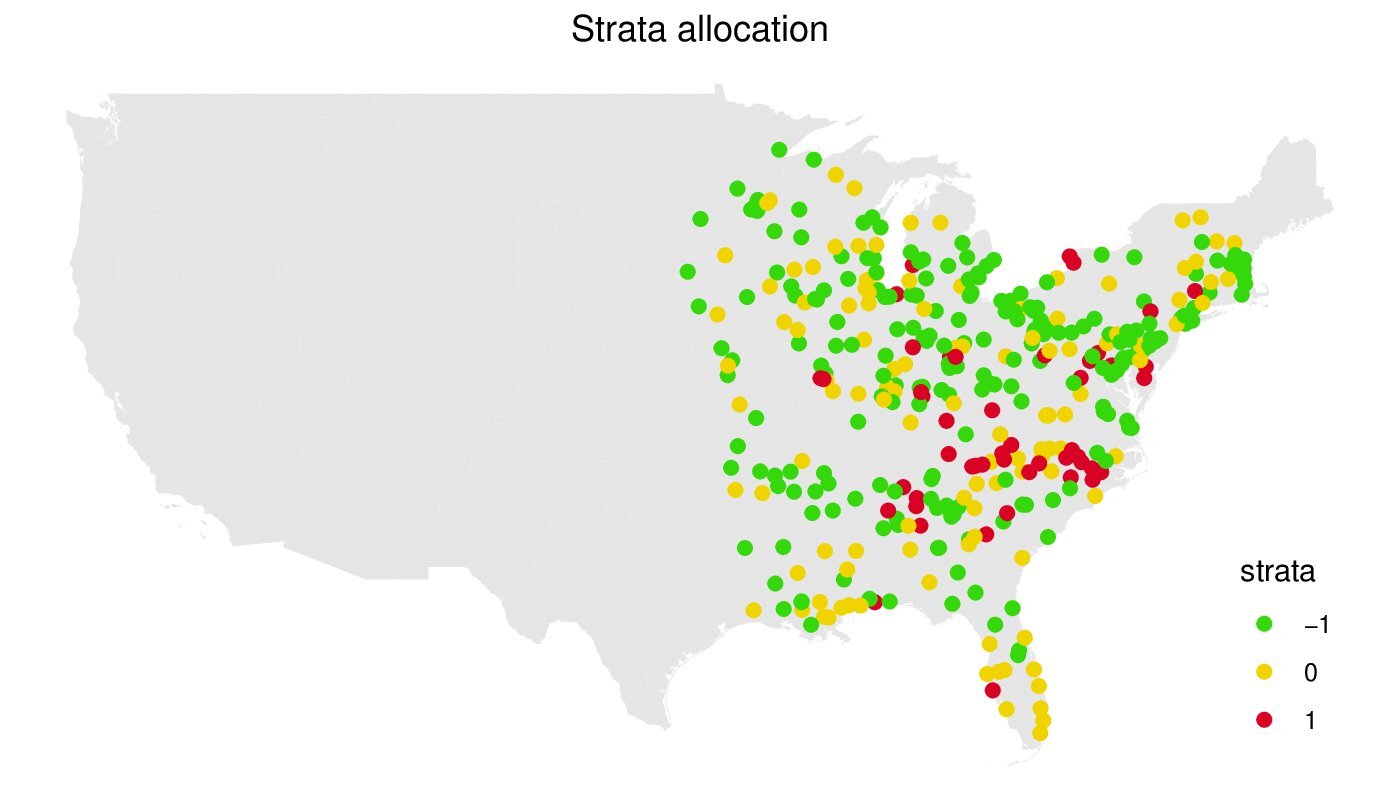}
\caption{Considered counties in the Easter U.S.. (top left) Probability to be allocated in the associative positive stratum. (top right) Probability to be allocated in the dissociative stratum. (bottom left) Probability to be allocated in the associative negative stratum. (bottom right) Point estimation of strata allocation.}
\label{fig:map_strata}
\end{figure}

In addition to the analysis reported in Section 5.2, our proposed approach allows us to quantify the uncertainty of strata allocation. In fact, for each county, we know the probability of being allocated in each of the three strata, in addition to the estimation of its allocation. In addition, we can visualize this information on the US map. Specifically, the first three maps in Figure \ref{fig:map_strata} visualize the probability that each county is assigned to the three different strata. As already underlined in Figure 3, %\ref{fig:covariates_P2}, 
counties with a higher probability of being allocated in the associative negative stratum and the dissociative stratum are far from the largest cities, different from the associative positive stratum. In addition, western countries seem to have a small probability of being assigned to the associative negative stratum. The fourth map, in bottom right in Figure \ref{fig:map_strata}, reports the estimation of the partition point of the strata, a partition that is used to estimate the principal causal effects.

\vspace{0.5cm}

%\begin{figure}[!htb]
%\begin{center}
%\includegraphics[trim={1cm 1.5cm 1cm 1.5cm}, width=2.5in]{Figures/application_aftercodem/spiderplot_3_new.pdf}\;\includegraphics[trim={1cm 1.5cm 1cm 1.5cm},width=2.5in]{Figures/application_aftercodem/spiderplot_2_new.pdf}\\ \includegraphics[trim={1cm 1.5cm 1cm 1.5cm},width=2.5in]{Figures/application_aftercodem/spiderplot_1_new.pdf}
%\end{center}
%\caption{Representation of the characteristics of the identified strata. Each spider plot reports in the colored area the strata-specific characteristics (the mean of the analyzed covariates) and in the gray area the collective characteristics (the mean of the covariates among all the analyzed counties in the Eastern U.S.). We can consider the gray area as the benchmark to understand how the characteristics of each stratum differ from the collective characteristics of the analyzed population. \label{fig:covariates_P2}}
%\end{figure}